\documentclass[twocolumn,showpacs,prd]{revtex4}
\usepackage{amsmath,amssymb,amsfonts,amsthm}

\usepackage{graphicx}

\newcommand{\bv}{\bar v}
\newcommand{\bbeta}{\beta^\rho}
\newcommand{\bV}{\bar V}

\newcommand{\Rt}{\mathbb{R}^3}

\newcommand{\Rdm}{\mathbb{R}^2_{+}}

\newcommand{\N}{\mathcal{N}}

\newcommand{\rc}{^{(4)}\mathcal R}

\newcommand{\rt}{{^{(3)}R}}

\newcommand{\rd}{{^{(2)}R}}

\newcommand{\ckq}{\mathcal{L}_q}
\newcommand{\ck}{\mathcal{L}}

\newcommand{\Lq}{\Delta_q}

\newcommand{\Ld}{\Delta}

\newcommand{\Ldt}{{^{(3)}\Delta}}

\begin{document}
\title{On well-posedness, linear perturbations and mass conservation 
for axisymmetric Einstein equation}

\author{Sergio Dain}
\affiliation{Facultad de Matem\'atica, Astronom\'{\i}a y F\'{i}sica, Universidad
Nacional de C\'ordoba, Ciudad Universitaria (5000) C\'ordoba, Argentina;}
\affiliation{Max Planck Institute for Gravitational Physics (Albert
  Einstein Institute) Am M\"uhlenberg 1 D-14476 Potsdam Germany;}
\affiliation{Instituto de F\'{\i}sica Enrique Gaviola, CONICET, UNC.}

\author{Omar E. Ortiz}
\affiliation{Facultad de Matem\'atica, Astronom\'{\i}a y F\'{i}sica, Universidad
Nacional de C\'ordoba, Ciudad Universitaria (5000) C\'ordoba, Argentina;}
\affiliation{Instituto de F\'{\i}sica Enrique Gaviola, CONICET, UNC.}

\date{\today}

\begin{abstract}
  For axially symmetric solutions of Einstein equations there exists a gauge
  which has the remarkable property that the total mass can be written as a
  conserved, positive definite, integral on the spacelike slices. The
  mass integral provides a nonlinear control of the variables along the whole
  evolution. In this gauge, Einstein equations reduce to a coupled
  hyperbolic-elliptic system which is formally singular at the axis.  As a
  first step in analyzing this system of equations we study linear
  perturbations on flat background. We prove that the linear equations reduce
  to a very simple system of equations which provide, thought the mass formula,
  useful insight into the structure of the full system. However, the singular
  behavior of the coefficients at the axis makes the study of this linear
  system difficult from the analytical point of view. In order to understand
  the behavior of the solutions, we study the numerical evolution of them. We
  provide strong numerical evidence that the system is well-posed and that its
  solutions have the expected behavior. Finally, this linear system allows us
  to formulate a model problem which is physically interesting by itself, since
  it is connected with the linear stability of black holes solutions in axial
  symmetry. This model can contribute significantly to solve the nonlinear
  problem and at the same time it appears to be tractable.
\end{abstract}

\maketitle

\section{Introduction}
\label{sec:introduction}

Axisymmetric spacetimes has been studied mainly for two reasons. The first one
is that they often appear in astrophysical models like rotating stars and black
holes.  The second is because in the presence of any symmetry Einstein
equations simplify considerable and hence these spacetimes are useful as
intermediate step to understand more complex problems.  In particular, axially
symmetric gravitational waves in vacuum do not carry angular momentum, this
represents an important simplification in the dynamics.  Also, axial symmetry is
the only symmetry compatible with asymptotic flatness and non-trivial
gravitational radiation \cite{bicak98}. From this perspective, axially
symmetric gravitational waves are the simplest possible waves emitted from
isolated sources. And hence they represent the natural candidates to study the
strong field dynamics of gravitational waves in Einstein equations.

However, axial symmetry presents a major difficulty. To take advantage of the
symmetry an adapted coordinate system should be used in order to reduce the
field equations to a lower-dimensional system (there is a well known procedure
to do this for any symmetry in a geometrical way \cite{Geroch71}, we review
this result in Sec. \ref{sec:symmetry-reduction}). The problem is that the
norm of the axial Killing vector vanishes at the axis, and hence the reduced
equations are formally singular there.   

This difficulty is so severe that until recently axially symmetric spacetimes
have not been studied in detail even using numerical techniques (see chapter
10.4 in \cite{alcubierre-it3nrsomop2008} and references therein).  In
a number of recent articles \cite{Garfinkle:2000hd}, \cite{Choptuik:2003as},
\cite{Rinne:2008tk}, \cite{Rinne:thesis}, \cite{Ruiz:2007rs} this kind of
singular behavior has been successfully implemented numerically.  There is
however no analytical study of axial symmetry in the dynamical regime (see the
review article \cite{Rendall:lrr-2005-6} for results for other kind of
symmetries). In fact, it can be argued that this singular behavior near the
axis is so complicated that the axially symmetric case is as hard as the full
general case from the analytical point of view.

There exists however a new ingredient that makes, in our opinion, the problem
worth studying.  In the article \cite{Dain:2008xr} it has been proved that
there exists a gauge in axial symmetry such that the total mass of the
spacetime can be written as a positive definite volume integral over the
spacelike slices of the foliation. Moreover, this integral is conserved along
the evolution. This conserved integral control the norm of the fields along the
whole evolution. This is certainly a very desirable property of this gauge
which is not present in the general, non-symmetric, case. Also, this mass
integral formula appears to be connected with stability properties of black
holes in axial symmetry \cite{Dain:2007pk}.

The gauge mentioned above is a combination of the well known maximal condition
for the lapse and the choice of isothermal coordinates (also called quasi
isotropical) for the shift. The later condition is only possible in axial
symmetry. We call it the maximal-isothermal gauge. This gauge has been known
for long time (see \cite{Bardeen83} and \cite{Nakamura87}) but without noticing
this property of the mass. It is also important to emphasize that this gauge is
the one used in most of the recent numerical computations
\cite{Garfinkle:2000hd} \cite{Rinne:thesis}
\cite{Rinne:2008tk}\cite{Choptuik:2003as} (examples of other gauge choices in
axial symmetry are given in \cite{Rinne:2005sk} \cite{Sorkin:2009wh}).  That
is, this gauge has not only desirable analytical properties but it is also
useful for numerical studies.

The very basic question of well-posedness of the equations in this gauge is
open. This question is rather subtle because of the singular behavior
mentioned above. The standard theory in partial differential equations does not
seems to apply in a direct way. This is the problem we want to study in this
article. In order to do this, the first step is to study the linearization of
the equation around fixed solutions. We chose Minkowski as a background for
simplicity. As we describe in the next section, we obtain a remarkable simple
system of linear equation together with a conserved quantity which corresponds
to the mass of the spacetime up to second-order corrections. This system allows
us to formulate the problem of well-posedness in a simplified setting which is
nevertheless relevant and physically interesting. Remarkably enough, even for
this linear system the well-posedness appears to be a nontrivial problem. In order
to get insight into this problem we numerically evolve these equations to provide
evidences that the system is in fact well-posed and that the solutions have the
expected behavior.
 
If the local existence problem is so complicated in this gauge one can wonder
what can be said about the global behavior of the evolution, which is, of
course, the ultimate goal. However, many of the main complications of this
gauge are already present in the well-posedness problem because they are related
with the local behavior of the fields at the symmetry axis. If one can solve them at
the linearized level in a satisfactory way there is a good chance that the mass
integral formula can be used to control the global evolution in some way. Also,
the well-posedness of the linear equations are relevant by themselves for the
following two reasons.  First, the
mass formula at the linear level can in principle be used to prove linear
stability in axial symmetry of a background solution like a black hole. Second,
the well-posedness of the linear equations and the mass formula give insight on
appropriate boundary conditions on a bounded domain. In particular, the mass
formula allows us to calculate the gravitational waves that leave or enter a bounded
domain.

The plan of the article is the following. In Sec. \ref{sec:main-results} we
summarize our mains results. In Sec. \ref{sec:axisymm-vacu-einst} we review
the axially symmetric, vacuum, Einstein equations. Although this is well
known, the way we process to obtain the final equations in the
maximal-isothermal gauge is slightly different than the standard one used in the
numerical works mentioned above.  In Sec. \ref{sec:linearized-equations} we
derive our main linear equations and in Sec. \ref{sec:prop-line-equat} we
describe their main properties. In particular, in this section we discuss
the mass conservation and boundary conditions on a bounded domain. In Sec.
\ref{sec:numerical-techniques} we describe the numerical techniques used to
evolve these equations. And in Sec. \ref{sec:numerical-results} we present
the numerical results. Finally, in Sec. \ref{sec:final-comments} we conclude
with a discussion of the relevant open problems.
  
\section{Main results}
\label{sec:main-results}

This article has two main results. The first one is to prove that the
linearized Einstein vacuum equations in the maximal-isothermal gauge reduce to a
very simple set of equations together with a conserved quantity. This conserved
quantity is the mass up to second-order corrections and it is written as 
a positive definite integral over a spacelike surface, which has a similar form
as the energy of the wave equation. This property of the mass, which only holds
in this gauge, is of course what distinguished this system of equations from any
other linearization. 

The second result is the numerical study of these equations, together with the
analysis of appropriate boundary conditions on a finite grid.

Let us describe the first result. In axial symmetry, the dynamical degrees of
freedom of the vacuum gravitational field are prescribed by two functions,
which can be chosen to be the norm and twist potential of the axial Killing
vector (see Sec. \ref{sec:axisymm-vacu-einst}). We make, for simplicity, the
extra assumption that the twist is zero (although we discuss the full
non-linear equations with twist in Sec. \ref{sec:axisymm-vacu-einst}). This
assumption simplify the equations but it is by no means essential. In the
maximal-isothermal gauge, the linearized Einstein equations with respect to a
Minkowski background reduce to the following two equations for the functions
$v$ and $\beta^\rho$ (the reason for the notation for the last function is that it
represents the $\rho$ component of the shift vector as we will see below)
\begin{align}
  \label{eq:119b}
  \ddot v  &= \Ld v  - \frac{\partial_\rho v }{\rho}+
  \rho \partial_\rho \left(\frac{\beta^\rho}{\rho}  \right),\\
  \label{eq:120b}
  \Ld \beta^\rho & = \frac{2}{\rho} \left(\Ld v - \frac{\partial_\rho
      v}{\rho}\right).  
\end{align}
These equations are deduced in Sec. \ref{sec:linearized-equations}.  We have
chosen cylindrical coordinates $(t,\rho,z)$. The relevant domain for these
equations is the half plane $\rho\geq 0$, $-\infty <z<\infty$ denoted by $\Rdm$.
A dot denotes time derivative and $\Ld$ is the flat Laplacian in 2-dimensions
\begin{equation}
  \label{eq:15}
\Ld v = \partial^2_\rho v +  \partial^2_z  v.
\end{equation}
The boundary condition for Eqs. \eqref{eq:119b} and \eqref{eq:120b} arise
from the regularity of the spacetime metric at the axis and the standard
asymptotically flat fall-off behavior at infinity. We discuss this in detail
in Sec. \ref{sec:bound-cond-axial} and \ref{sec:prop-line-equat}. Let us
present here a summary.  Eq. \eqref{eq:120b} is an elliptic equation for
$\beta^\rho$, we need to prescribe boundary conditions on $\Rdm$. On the axis
$\rho=0$ we require
\begin{equation}
  \label{eq:84}
\beta^\rho|_{\rho=0} =0, 
\end{equation}
and at infinity we impose
\begin{equation}
  \label{eq:146}
 \beta^\rho =O(r^{-1}),  
\end{equation}
where $r=\sqrt{\rho^2+z^2}$.
With these boundary conditions, Eq. \eqref{eq:120b} has a unique
solution. Eq. \eqref{eq:119b} is a wave equation for $v$, we need to
prescribe initial conditions, which are functions $f(z,\rho)$ and $g(z,\rho)$
such that
\begin{equation}
  \label{eq:147}
  v|_{t=0}=f, \quad  \dot v|_{t=0}=g.
\end{equation}
The axis  represents a timelike boundary for the wave equation
\eqref{eq:119b}, and hence we need to prescribe also boundary conditions there. 
This is the delicate part, because the equations are  singular at the
axis and hence we are not free to chose arbitrary boundary conditions
there. From the axial regularity of the spacetime metric we deduce that  the
initial data $f$ and $g$ should vanish at the axis, namely
\begin{equation}
  \label{eq:25}
  f|_{\rho=0}=0, \quad  g|_{\rho=0}=0. 
\end{equation}
In Sec. \ref{sec:prop-line-equat}, using series expansions, we prove
that conditions \eqref{eq:25} on the initial data imply that
\begin{equation}
  \label{eq:148}
  v|_{\rho=0}=0, \quad \partial_\rho v|_{\rho=0}=0,
\end{equation}
for all times.  Moreover, solutions $v$ and $\beta^\rho$ of equations
\eqref{eq:119b} and \eqref{eq:120b} satisfy a parity conditions, namely $v$ is an
even function of $\rho$ and $\beta^\rho$ is an odd function of $\rho$. These
parity conditions imply that the spacetime metric is smooth at the axis.  It is
important to emphasize that these conditions  are consequences of  equations
\eqref{eq:119b} and \eqref{eq:120b} alone, without any extra requirement.

In the numerical implementation, Eqs. \eqref{eq:148} are used as boundary
conditions at the axis.  There are various ways to re-express
\eqref{eq:119b} and \eqref{eq:120b} in order write conditions \eqref{eq:148} as
proper timelike boundary  conditions (e.g. Dirichlet or
Neumann). For example, following \cite{Garfinkle:2000hd}, in Sec.
\ref{sec:numerical-techniques} we write them in terms of the rescaled variable
$\bar v = v/\rho$. 

We are interested in asymptotically flat solutions of
\eqref{eq:119b} and \eqref{eq:120b}. 
We will argue in Sec. \ref{sec:prop-line-equat}, that the typical fall off behavior
as $r\to \infty$ for this kind of solutions is
\begin{equation}
  \label{eq:29}
  v=O(r^{-2}).
\end{equation}
That is, if we chose initial data $f$ and $g$ which satisfy \eqref{eq:29} then
$v$ will satisfy \eqref{eq:29} for all times.

All the other components of the linear perturbation can be calculated in terms
of $v$ and $\beta^\rho$ as follows.  In our gauge the four dimensional
coordinates are given by $(t,\rho,z,\phi)$. A general twist free linear
perturbation is written as follows 
\begin{equation}
  \label{eq:106}
 \gamma =  (\sigma+2q)(d\rho^2+dz^2)+2\beta^\rho d\rho dt +2\beta^z dzdt +
 \rho^2\sigma d\phi^2.          
\end{equation}
Where the functions $\sigma$, $q$, $\beta^\rho$ and $\beta^z$ depends only on
$(t,\rho,z)$.  The function $\beta^\rho$ is given by (\ref{eq:120b}), the other
functions are calculated in terms of $v$ as follows. The functions $q$ 
is a time derivative of $v$
\begin{equation}
  \label{eq:109}
  q=\dot v.
\end{equation} 
The function
$\sigma$ is  determined  by the following elliptic equation 
\begin{equation}
  \label{eq:86}
  \Ldt \sigma =-\Ld \dot v, 
\end{equation}
where $\Ldt$ is defined as 
\begin{equation}
  \label{eq:121}
  \Ldt \sigma =\Ld\sigma +\frac{\partial_\rho \sigma}{\rho}.
\end{equation}
This operator, which appears frequently in the rest of the article, is the flat
Laplace operator in 3-dimensions written in cylindrical coordinates and acting
on axially symmetric functions. The boundary condition for equation
(\ref{eq:86}) at the axis is given by
\begin{equation}
  \label{eq:125}
  \partial_\rho \sigma|_{\rho=0} =0, 
\end{equation}
and  at infinity we impose
\begin{equation}
  \label{eq:129}
  \sigma =O(r^{-1}).
\end{equation}
Eq. \eqref{eq:86} can be also viewed as an equation in $\Rt$. In this case
we do not need to prescribe any boundary condition at the axis. Condition
\eqref{eq:125} will be automatically satisfied for any regular solution. 

Finally, the other component of the shift vector is determined by the following equation
\begin{equation}
  \label{eq:110}
   \Ld \beta^z= -2 \frac{\partial_z v }{\rho^2},
\end{equation}
with boundary condition at the axis
\begin{equation}
  \label{eq:130}
   \partial_\rho \beta^z|_{\rho=0} =0,
\end{equation}
and decay condition at infinity
\begin{equation}
  \label{eq:131}
 \beta^z =O(r^{-1}). 
\end{equation}

The total mass of the system is given by the following integral
\begin{equation}
  \label{eq:124}
  m=\frac{1}{16}\int_{\Rdm} \left(4\frac{|\partial v|^2}{\rho^2}+(\Ld v)^2 +
    |\partial \sigma |^2 \right)\rho \, d\rho dz.
\end{equation} 
Note that in order to compute the mass we need the function $\sigma$, which
satisfies Eq. \eqref{eq:86}. This equation is uncoupled with equations
\eqref{eq:119b} and \eqref{eq:120b}.  The integral \eqref{eq:124} is
conserved. That is, for every solution of \eqref{eq:119b} and \eqref{eq:120b}
which satisfy the boundary conditions \eqref{eq:84}, \eqref{eq:146},
\eqref{eq:148} and decay at infinity like \eqref{eq:29} we have
\begin{equation}
  \label{eq:112}
  \dot m=0. 
\end{equation}
The conservation law \eqref{eq:112} is deduced from a local conservation
formula which involves the  integrand of the mass formula \eqref{eq:124}. This
local conservation law can be also used to compute the gravitational waves
entering or leaving  a bounded domain. We discuss this in
Sec. \ref{sec:prop-line-equat}.

The second main result of this article is the numerical study of the system
\eqref{eq:119b} and \eqref{eq:120b}. We describe this in detail in Sec.
\ref{sec:numerical-results}. Let us briefly summarize these results.  The system
\eqref{eq:119b} and \eqref{eq:120b} appears, from the numerical evidences, to be
well posed and numerically stable. In particular, this imply that the functions
$v$ and $\beta^\rho$ remain bounded for all times by a constant that depends
only on the initial data. This is consistent with the linear stability of
Minkowski spacetime.

The numerical calculations are, of course, performed on a finite grid. Hence, we
need to prescribe boundary conditions on a bounded domain. These conditions
should be compatible with asymptotically flatness in the following sense. Assume
we have a sequence of bounded domains such that in the limit they cover the
half plane $\Rdm$.  If we solved the equations for this sequence of domains we
should recover in the limit the asymptotically flat solution described
above. There exists many different boundary conditions that have this
property. In particular, homogeneous Dirichlet conditions for $\beta^\rho$ and
$v$. For each bounded domain the mass is not conserved. However, as the size of
the domain increase we expect that the mass approach a time independent
constant. This is precisely what we observe in our numerical calculations.  

For our present goal, this kind of asymptotically flat boundary conditions is
all what we need. There is, however, an interesting extra point here. To model
an isolated system on a finite grid it is important to prescribe boundary
conditions such that the gravitational radiation leaves the domain. In general,
this is a very difficult problem since it is not even clear what we mean by
gravitational radiation at a finite distance. However, as we mention above, in
our gauge the mass formula allow us to compute gravitational radiation on a
bounded domain. Although it appears not to be possible to prescribe boundary
conditions such that the gravitational waves always leave the domain, the mass
formula suggests a particular kind of boundary conditions that has this
behavior in our numerical calculations.  That is, under these boundary
conditions, the mass on a bounded domain is monotonically decreasing with time
for the particular kind of initial data used in the computations.  We emphasize
however that we have not been able to prove this analytically.  We explore this
in detail in Sec. \ref{sec:prop-line-equat} and \ref{sec:numerical-results}.

\section{Axisymmetric vacuum Einstein equations}
\label{sec:axisymm-vacu-einst}

The purpose of this section is to write the vacuum Einstein equations for
axially symmetric spacetimes in the maximal-isothermal gauge. This involves
three clearly distinguished steps. In the first one, described in Sec.
\ref{sec:symmetry-reduction}, we perform a symmetry reduction of Einstein
equations to obtain a set of geometrical equations in the 3-dimensional
quotient manifold. These equations can be viewed as 3-dimensional Einstein
equations coupled with effective matter sources. In the second step (Sec.
\ref{sec:2+1-decomposition}) we chose an arbitrary spacelike foliation in the quotient
manifold and split the equations in time plus space. In Sec. \ref{sec:gauge}
we fixes the foliation and the coordinate system. We also write the mass formula
in this gauge. Finally, in Sec. \ref{sec:bound-cond-axial} we discuss
boundary conditions at the axis and at infinity.

\subsection{Symmetry reduction}
\label{sec:symmetry-reduction}

In this section we perform the symmetry reduction of the field equations. 
We follow \cite{Geroch71} and \cite{Weinstein90}. See also
\cite{Choptuik:2003as}  \cite{Rinne:2005sk}.  

Consider a vacuum solution of Einstein's equations, i.e., a four dimensional
manifold $M$ with metric $g_{\mu\nu}$ (with signature $(-+++)$) such that the
corresponding Ricci tensor vanishes
\begin{equation}
  \label{eq:132}
  \rc_{\mu\nu}=0. 
\end{equation}
Suppose, in addition, that the metric $g_{\mu\nu}$ admits  a
Killing field  $\eta^\mu$, that is $\eta^\mu$  satisfies the equation
\begin{equation}
  \label{eq:95}
  \hat \nabla_{(\mu} \eta_{\nu)} =0,
\end{equation}
where $\hat \nabla_\mu$ is the connection with respect to $g_{\mu\nu}$.  Greek
indices $\mu, \nu, \cdots $ denote  four dimensional indices. 

We define the square of the norm and the
twist of $\eta^\mu$, respectively, by
\begin{equation}
  \label{eq:1}
\eta =\eta^\mu\eta^\nu g_{\mu\nu}, \quad  \omega_\mu=\epsilon_{\mu\nu\lambda
  \gamma }\eta^\nu \hat
\nabla^\lambda\eta^\gamma.
\end{equation}
Using the field Eq. \eqref{eq:132}  it is possible to prove that 
\begin{equation}
  \label{eq:96}
  \hat \nabla_{[\mu} \omega_{\nu]}=0,
\end{equation}
and hence $\omega_\mu$ is locally the gradient of a
scalar field $\omega$
\begin{equation}
  \label{eq:2}
 \omega_\mu=\hat \nabla_\mu \omega. 
\end{equation}

Let $\mathcal{N}$ denote the collection of all trajectories of $\eta^\mu$, and
assume that it is a differential 3-manifold.
We define the metric $h_{\mu\nu}$  on $\mathcal{N}$ by
\begin{equation}
  \label{eq:3}
\eta g_{\mu\nu}=h_{\mu\nu}+ \eta_\mu\eta_\nu.
\end{equation}
 The vacuum field equations
\eqref{eq:132} can be written in the following form on $\mathcal{N}$
\begin{align}
  \label{eq:4}
 \Box \eta & =\frac{1}{\eta}(\nabla^a \eta \nabla_a \eta
- \nabla^a \omega \nabla_a \omega), \\
 \Box \omega & =\frac{2}{\eta}\nabla^a\omega \nabla_a \eta,\label{eq:4b} \\
\rt_{ab} &= \frac{1}{2\eta^2} (\nabla_a \eta\nabla_b \eta + \nabla_a \omega \nabla_b
\omega).  \label{eq:4c} 
\end{align}
where $\nabla_a$ and $\rt_{ab}$ are the connexion and the Ricci tensor of
$h_{ab}$, we have defined $\Box =\nabla_a \nabla^a$ and Latin indices
$a,b\ldots$ denote three dimensional indices on $\mathcal{N}$.

Note that the definition of the metric \eqref{eq:3} involves a conformal
rescaling with respect to the canonical
metric  $\tilde h_{\nu\mu}$  defined by  
\begin{equation}
  \label{eq:16}
g_{\nu\mu}= \tilde h_{\nu\mu}+ \eta^{-1}\eta_\nu\eta_\mu.  
\end{equation}
That is, we have
\begin{equation}
  \label{eq:17}
  h_{\mu\nu}=\eta\tilde h_{\mu\nu}.
\end{equation}
This rescaling simplify considerably the field equations.  In particular, on
the right hand side of Eq. \eqref{eq:4c} there are no second derivatives
of the fields $\eta$ and $\omega$ (compare, for example, with equation (20) in
\cite{Dain:2008xr}).

Finally, we note that Eq. (\ref{eq:4}) can be written in the following
form
\begin{equation}
  \label{eq:85}
 \Box \Sigma = - \frac{\nabla^a \omega \nabla_a \omega}{\eta^2}, 
\end{equation}
where we have defined   
\begin{equation}
  \label{eq:97}
  \Sigma = \log \eta. 
\end{equation}

Up to this point, the only assumption we have made is that the spacetime admits
a Killing vector field $\eta^\mu$ and that $\eta^\mu$ is not null, otherwise 
the metric $h_{ab}$ is not defined. If the Killing field is timelike ($\eta<0$)
then the metric $h_{ab}$ is Riemannian and the equations
\eqref{eq:4}--\eqref{eq:4c} are the stationary Einstein vacuum equations.  On
the other hand, when the Killing vector is spacelike ($\eta>0$), the metric
$h_{ab}$ is a is a 3-dimensional Lorenzian metric (we chose the signature
$(-++)$). In axially symmetry, the Killing vector $\eta^\mu$ is spacelike and
its norm vanishes at the axis of symmetry. Hence, the equations are formally
singular at the axis. This singular behavior at the axis represents the main
difficulty to handle these equations.

In the Lorenzian case, 
Eq. \eqref{eq:4c} has the form of Einstein equations in three dimensions,
with effective matter sources produces by $\eta$ and $\omega$. The effective
matter Eqs. \eqref{eq:4}--\eqref{eq:4b} imply that the energy-momentum
tensor defined in terms of $\eta$ and $\omega$ by 
\begin{multline}
  \label{eq:107}
  T_{ab}=\frac{1}{2\eta^2} (\nabla_a \eta\nabla_b \eta + \nabla_a \omega \nabla_b
\omega)-\\
\frac{1}{4\eta^2} h_{ab}(\nabla_c \eta\nabla^c \eta + \nabla_c \omega \nabla^c
\omega),
\end{multline}
is divergence free, i.e. $\nabla^aT_{ab}=0$. 

A particularly relevant special case  is when $\omega=0$. In
that case Eqs. \eqref{eq:4}--\eqref{eq:4b} simplify considerable
\begin{align}
  \label{eq:tf1}
\Box \Sigma & =0, \\
\rt_{ab} &= \frac{1}{2} \nabla_a \Sigma \nabla_b \Sigma.  \label{eq:tf2} 
\end{align}

We have pointed out that the rescaling \eqref{eq:17} simplifies the equations and
allow us to write them in a more geometric form.  This is the reason why this
scaling is used in the case of $U(1)$ cosmologies where the equations are
locally the same but the norm $\eta$ never vanishes (see
\cite{Choquet-Bruhat01} \cite{Choquet-Bruhat04} and the review article
\cite{andersson04}). In our case the conformal scaling \eqref{eq:17} is
singular at the axis. However, since the behavior of $\eta$ at the axis, as we
will see in the next sections, is controlled a priori this singular scaling does
not seems to introduce any extra difficulty in the equations.  We also remark
that in all the numerical works mentioned above this conformal rescaling was not
used, the equations are written in terms of the metric $\tilde h_{ab}$
defined by \eqref{eq:16}. 

Eqs. (\ref{eq:4})--(\ref{eq:4c}) are purely geometric with respect to the
metric $h_{ab}$.  To solve these equations we need to prescribe some
gauge for the metric $h_{ab}$. This will be done in the next two sections.  

\subsection{2+1 decomposition}
\label{sec:2+1-decomposition}
In order to formulate an initial value problem, we will perform an standard
$2+1$ decomposition of Eqs. (\ref{eq:4})--(\ref{eq:4c}). Note that this is
completely analogous to the $3+1$ decomposition of Einstein equations, in fact
all the formulas are formally identical because the dimension do not appears
explicitly in them (see, for example, \cite{rendall-pdeigrgtim2008},
\cite{rendall08}).
  
Consider a foliation of spacelike, 2-dimensional slices $S$ of the metric
$h_{ab}$. Let $t$ be an associated time function  and let $n^a$ be
the unit normal vector orthogonal to $S$ with respect to the metric
$h_{ab}$. The intrinsic metric on $S$ is denoted by $q_{ab}$ and is given by
\begin{equation}
  \label{eq:6}
h_{ab}=-n_an_b + q_{ab}.
\end{equation}
Define the density $\mu$ by
\begin{equation}
  \label{eq:5}
\mu = 2\rt_{ab} n^a n^b+ \rt,
\end{equation}
 and the current $J_b$ by
\begin{equation}
  \label{eq:36}
  J_b=-q^c_bn^a \rt_{ca},
\end{equation}
where  $\rt=\rt_{ab}h^{ab}$  denotes the trace of $\rt_{ab}$.
 Then, using Eq. \eqref{eq:4c} we obtain
\begin{align}
  \label{eq:7}
\mu &= \frac{1}{2\eta^2}\left(\eta'^2 + \omega'^2 + |D\eta|^2 + |D\omega|^2
\right),\\
J_A &= -\frac{1}{2\eta^2}\left(\eta' D_A\eta  +  \omega' D_A\omega \right),
\end{align}
where $D_A$ is the connexion with respect to $q_{AB}$. The prime denotes
directional derivative with respect to $n^a$, that is
\begin{equation}
  \label{eq:14}
\eta' = n^a\nabla_a \eta= \frac{1}{\alpha} \left(\partial_t\eta- \beta^AD_A
  \eta\right) 
\end{equation}
where $\alpha$ is the lapse and $\beta^A$ is the shift vector of the
foliation.
The indices $A,B, \cdots$ denotes two dimensional indices on $S$. 
The constraints equations corresponding to (\ref{eq:4c}) are  given by 
\begin{align}
  \label{eq:8}
\rd  -\chi^{AB}\chi_{AB}+\chi^2 &=\mu,\\
 D^A\chi_{AB} -D_B\chi &=J_B, \label{eq:8b}
\end{align}
where $\rd$ is the Ricci scalar of $q_{AB}$,  $\chi_{AB}$ is the
second fundamental form of $S$ and $\chi$ its trace
\begin{equation}
  \label{eq:133}
  \chi= q^{AB}\chi_{AB}. 
\end{equation}
We use the following sign convention for the definition of $\chi_{AB}$
\begin{equation}
  \label{eq:sce}
  \chi_{ab}= - q^c_a \nabla_c n_b= -\frac{1}{2}\pounds_n q_{ab},
\end{equation}
where  $\pounds$ denotes Lie derivative.
The  evolution equations are given by
\begin{align}
  \label{eq:127}
 \partial_t q_{AB} &= -2\alpha \chi_{AB} + \pounds_\beta q_{AB},\\ 
\label{eq:38}
\partial_t \chi_{AB} & = \pounds_\beta \chi_{AB}- D_AD_B\alpha +\alpha \tau_{AB},
\end{align}
where
\begin{equation}
  \label{eq:134}
  \tau_{AB}=\chi \chi_{AB}  + \rd_{AB}-\rt_{AB}-2\chi_{AC}\chi^{C}_{B}.
\end{equation}
and  
\begin{equation}
  \label{eq:136}
  \rt_{AB}=\frac{1}{2\eta^2} (\partial_A \eta \partial_B \eta + \partial_A
  \omega \partial_B \omega).  
\end{equation}
The evolution equations
\eqref{eq:127}--\eqref{eq:38} and the constraint equations
\eqref{eq:8}--\eqref{eq:8b} constitute a complete $2+1$ decomposition of the
3-dimensional Einstein Eq. \eqref{eq:4c}. It remains to decompose the
effective matter Eqs. \eqref{eq:4}--\eqref{eq:4b}. This can easily be
obtained using the decomposition formula \eqref{eq:68} for the wave operator
$\Box$ and the definition of the
metric $q_{ab}$ given by \eqref{eq:6}. The result is the following
\begin{align}
  \label{eq:149}
  -\Sigma'' + \Lq \Sigma + D_A \Sigma \frac{D^A \alpha}{\alpha} + \Sigma' \chi
  & =\frac{1}{\eta^2} \left(\omega'^2 -|D\omega|^2 \right),\\
 \label{eq:149b}
-\omega'' + \Lq \omega + D_A \omega \frac{D^A \alpha}{\alpha} + \omega' \chi 
&= \frac{2}{\eta^2}\left(D_A\omega D^A\eta -\omega'\eta'\right),
\end{align}
where instead of \eqref{eq:4} we have use \eqref{eq:85}, and  $\Lq$ is the
Laplacian with respect to $q_{AB}$, i.e.  $\Lq=D^AD_A$.

Finally, we mention that the line element  of the metric $h_{ab}$  takes the
standard form
\begin{equation}
  \label{eq:26}
h = -\alpha^2dt^2+ q_{AB}(dx^A+\beta^A dt)(dx^B+\beta^B dt).
\end{equation}

\subsection{Gauge}
\label{sec:gauge}
In this section we describe the maximal-isothermal gauge. In particular we
review the mass formula for this gauge (see \cite{Dain:2008xr} for details).  For the
lapse, we impose the maximal condition on the 2-surfaces
\begin{equation}
  \label{eq:40}
  \chi=0.
\end{equation}
Note that we are not imposing that the surfaces are maximal in the
3-dimensional picture as in \cite{Dain:2008xr}. The later condition is the one
generally used \cite{Choptuik:2003as} \cite{Rinne:2008tk}, but the difference is
only minor. In particular the mass formula is positive definite for both
conditions as we will see. The one used here appears to be natural with respect
to the rescaled metric $h_{ab}$. Eq. \eqref{eq:40} implies the following
well known equation for the lapse
\begin{equation}
  \label{eq:37}
  \Lq \alpha = \alpha(\chi^{AB}\chi_{AB}+ \mu_1 ),
\end{equation}
where 
\begin{equation}
  \label{eq:45}
 \mu_1= \rt_{ab}n^an^b=  \frac{1}{2\eta^2}\left(\eta'^2 + \omega'^2 \right).
\end{equation}
The maximal gauge \eqref{eq:40} can be, of course, imposed in any dimensions
and  it is not related at all with axial symmetry. In contrast, the
condition for the shift is peculiar for two space dimensions.   
The shift vector is fixed by the requirement that 
the intrinsic metric $q_{AB}$ has the following form 
\begin{equation}
  \label{eq:10}
q_{AB}=e^{2u}\delta_{AB},
\end{equation}
where $\delta_{AB}$ is a fixed (i.e. $\partial_t \delta_{AB}=0$) flat metric in
two dimensions.  Then, using \eqref{eq:40}, we obtain that the trace free part
of \eqref{eq:127} is given by
\begin{equation}
  \label{eq:42}
  2\alpha\chi_{AB}=(\ckq \beta )_{AB},
\end{equation}
where $\ckq$ is the conformal Killing operator in two dimensions with respect
to the metric $q_{AB}$ defined in Eq. \eqref{eq:43}.  Equation
(\ref{eq:42}) is an elliptic first order system of equations for $\beta^A$.

 The elliptic Eqs. (\ref{eq:37}) and (\ref{eq:42}) determine lapse and
 shift for the metric $h_{ab}$ and hence fixes completely the gauge freedom in
 Eqs. \eqref{eq:4}--\eqref{eq:4c}. This gauge has associate a natural
 cylindrical coordinate system $(t, \rho, z)$ for which the metric
 $\delta_{AB}$ is given 
\begin{equation}
  \label{eq:66}
  \delta = d\rho^2+dz^2, 
\end{equation}  
and the axis of symmetry is given by $\rho=0$. The slices $S$ are the half
planes $\Rdm$. 

For the analysis of the equations it is of course important to write them explicitly as
partial differential equations in these coordinates.  We will do this in the
remainder of this section. In general, due to the complexity of Einstein
equations, the partial differential equations obtained in a particular gauge
can be quite involved. In our case, however, the geometric nature of the gauge
plus the symmetry reductions will provide a relative simple set of  equations. 

We first present some useful definitions. We need to subtract from $\eta$ the
part that vanishes at the axis. We define the function $\sigma$ by
\begin{equation}
  \label{eq:72}
  \eta=\rho^2 e^\sigma.
\end{equation}
Due to the rescaling \eqref{eq:17}, the lapse $\alpha$ vanishes also at the
axis, hence we define  the normalized lapse $\bar \alpha$ by
\begin{equation}
  \label{eq:71}
  \alpha=\rho \bar \alpha. 
\end{equation}
From the regularity conditions presented in the next section we will see that
it is useful to define the function $q$ defined by
\begin{equation}
  \label{eq:70}
  u=\log \rho + \sigma + q.
\end{equation}

We now proceed to write the equations.
We begin with the evolution equations for  $\sigma$ and $\omega$. The evolution
equation for $\sigma$ is given by 
\eqref{eq:149}. Using the definition \eqref{eq:72} and the conformal rescaling
expression for the Laplacian \eqref{eq:64} we obtain  
\begin{multline}
  \label{eq:137}
 -e^{2u} \sigma'' + \Ldt \sigma + \partial_A \sigma \frac{\partial^A \bar
   \alpha}{\bar \alpha}-2e^{2u} (\log\rho)'' +
 2\frac{\partial_\rho\bar\alpha}{\bar \alpha\rho}= \\
\left(-e^{2u}(\omega')^2+|\partial \omega|^2 \right)\rho^{-4}e^{-2\sigma}.  
\end{multline}
In the same way, from \eqref{eq:149b} we get 
\begin{multline}
  \label{eq:138}
  -e^{2u} \omega'' + \Ldt \omega + \partial_A \omega \frac{\partial^A \bar
    \alpha}{\bar \alpha}=\\
  \frac{2}{\eta}\left(-e^{2u}\omega'\eta'+ \partial_A\omega \partial^A\eta\right).
\end{multline}
Where $\partial_A$ denotes partial derivatives with respect to $\rho$ and $z$
and all the indices are moved with respect to the flat metric $\delta_{AB}$. In
these equations the lapse $\alpha$ and the shift $\beta^A$ appear trough the
prime operator defined in \eqref{eq:14}.

The momentum constraint (\ref{eq:8b}) is given by
\begin{equation}
  \label{eq:65}
  \partial^B \chi_{AB}= \bar J_A,
\end{equation}
where $\bar J_A=e^{2u}J_A$, that is we have
\begin{equation}
  \label{eq:98}
 \bar J_A = -\frac{e^{2u}}{2\eta^2}\left(\eta' \partial_A\eta  +
   \omega' \partial_A\omega \right). 
\end{equation}
To obtain \eqref{eq:65} we have used the conformal rescaling of the divergence
in 2-dimensions given by \eqref{eq:152}. The indices in Eq. \eqref{eq:65}
and in the rest of the article, are moved with the flat metric
$\delta_{AB}$. To avoid confusion, it is useful to introduce the following notation 
\begin{equation}
  \label{eq:153}
  \hat \beta_A = \beta^A\delta_{AB}, \quad \hat \chi^A_B= \delta^{AC}\chi_{CB}, \quad
  \hat \chi^{AB}= \delta^{AC}  \delta^{BD} 
  \chi_{CD}. 
\end{equation}
That is, we want to distinguish between, say, the covector $\beta_A =
\beta^Aq_{AB}$ used in the previous section and $ \hat \beta_A$ (see the
discussion after Eq. \eqref{eq:45b} in the Appendix).

The Hamiltonian constraint, Eq. (\ref{eq:8}), is given by 
\begin{equation}
  \label{eq:101}
  \Ldt \sigma +\Ld q = -\frac{\epsilon}{4},
\end{equation}
where
\begin{equation}
  \label{eq:102}
 \epsilon =   \frac{e^{2u}}{\eta^2}\left(\eta'^2 + \omega'^2 \right)+ |\partial
 \sigma|^2 + \frac{|\partial \omega|^2}{\eta^2} +  2 e^{-2u}  \hat\chi^{AB} \chi_{AB}. 
\end{equation}

Let us consider the evolution equations for $q_{AB}$ and $\chi_{AB}$. 
The evolution equation for the metric $q_{AB}$ reduces to 
\begin{equation}
  \label{eq:164}
  2\partial_tu=\partial_A\beta^A +2 \beta^A \partial_A u. 
\end{equation}
And the evolution equation for the second fundamental form $\chi_{AB}$ is given
by 
\begin{equation}
  \label{eq:140}
 \partial_t \chi_{AB}  = \pounds_\beta \chi_{AB}- F_{AB} -\alpha G_{AB} 
 -2\alpha \chi_{AC}\hat\chi^{C}_{B}  
\end{equation}
where $F_{AB}$ denotes the trace free part (with respect to $\delta_{AB}$) of 
$D_AD_B\alpha$. Using Eq. \eqref{eq:87}) we obtain
\begin{equation}
  \label{eq:154}
  F_{AB}=  \partial_A\partial_B\alpha -\frac{1}{2}\delta_{AB}\Ld\alpha-2\partial_{(A}\alpha \partial_{B)}u
 + \partial_C\alpha \partial^Cu \delta_{AB}.  
\end{equation}
And $G_{AB}$ denotes the trace free part of $\rt_{AB}$, namely
\begin{equation}
  \label{eq:155}
   G_{AB}= \rt_{AB}- \frac{1}{2}\delta_{AB}\rt_{CD} \delta^{CD}, 
\end{equation}
where $\rt_{AB}$ is given by \eqref{eq:136}. 

The equation for the lapse is given by 
\begin{equation}
  \label{eq:49}
   \Ld \alpha = \alpha  \left (e^{-2u} \hat\chi^{AB}  \chi_{AB} +e^{2u}
     \mu_1 \right ), 
\end{equation}
and for the shift we have
\begin{equation}
  \label{eq:142}
(\ck\beta )^{AB}=   2\alpha e^{-2u}  \hat\chi^{AB},
\end{equation}
where $\ck$ is the flat conformal Killing operator defined by \eqref{eq:67}.

Using the identity \eqref{eq:28}, we can transform the the first order system
of Eqs. \eqref{eq:142} for the shift and for the momentum constraint
\eqref{eq:65} in a pair of second-order uncoupled equations.  For
the shift, we take a divergence to Eq. \eqref{eq:142} to obtain
\begin{equation}
  \label{eq:50}
   \Ld \beta^A=2\partial_B (\alpha  \hat\chi^{AB}e^{-2u} ).
\end{equation}
For Eq. \eqref{eq:65} we define the vector $v^A$ by 
\begin{equation}
  \label{eq:46}
  \chi_{AB}=\ck (v)_{AB},
\end{equation}
and hence Eq. \eqref{eq:142} transform to 
\begin{equation}
  \label{eq:47}
  \Ld v_A= \bar J_A.
\end{equation}

The total ADM mass of the spacetime can be calculated as a volume integral on
the half plane $\Rdm$ of the positive definite effective energy density
\eqref{eq:102} (see \cite{Dain:2008xr})
\begin{equation}
  \label{eq:103}
  m= \frac{1}{16}\int_{\Rdm}   \epsilon\, \rho \,
  d\rho dz.   
\end{equation}
 
Finally, we mention that for the twist free case ($\omega=0$) the four
dimensional spacetime metric $g_{\mu\nu}$ has a simple expression in these
coordinates, namely
\begin{multline}
  \label{eq:128}
 g = -\frac{\alpha^2}{\rho^2}e^{-\sigma} dt^2+ \\
e^{\sigma+2q}\left((d\rho+\beta^\rho dt)^2+ (dz+\beta^z
  dt)^2\right)+\rho^2e^\sigma d\phi^2. 
\end{multline}

\subsection{Boundary conditions and axial regularity}
\label{sec:bound-cond-axial}
The boundary conditions at the axis in axial symmetry have been extensible
analyzed in the literature \cite{Garfinkle:2000hd}, \cite{Rinne:2005sk},
\cite{Rinne:thesis}, \cite{Choptuik:2003as}. They involve parity conditions in
the $\rho$ dependence of the different fields. That it, the relevant functions
are either even or odd functions of $\rho$. In order to use these results in
our setting, it is useful to write the relations of the quantities with respect
to the rescaled metric $h_{ab}$ and the canonical metric $\tilde h_{ab}$, since
all the above mentioned articles work with the metric $\tilde h_{ab}$.

Using relation \eqref{eq:17} we obtain for the 2-dimensional metric
\begin{equation}
  \label{eq:18}
  q_{AB}=\eta \tilde q_{AB},
\end{equation}
and for the second fundamental form
\begin{equation}
  \label{eq:13}
   \chi_{AB}= \sqrt{\eta} \left(\tilde
     \chi_{AB}+\frac{1}{2}\frac{\eta'}{\eta}\tilde q_{AB} \right), 
\end{equation}
where quantities with a tilde are written with respect to the metric $\tilde
h_{ab}$.  
We also have 
\begin{equation}
  \label{eq:88}
  \alpha= \sqrt{\eta} \tilde \alpha, \quad \beta^A= \tilde \beta^A.
\end{equation}
Using these relations and the results mentioned above it is straightforward to
obtain the following behavior of the relevant variables
\begin{equation}
  \label{eq:10c}
\eta,  \omega, \bar\alpha,  u, q, \sigma, \chi_{\rho\rho}, \beta^z \text{ are
  even functions of }\rho, 
\end{equation}
and
\begin{equation}
  \label{eq:9c}
  \chi_{\rho z}, \beta^\rho \text{ are odd functions of }\rho.
\end{equation}
Note that odd functions vanishes at the axis and the $\rho$ derivative of even
functions vanishes at the axis. It follows that one can impose homogeneous
Dirichlet boundary conditions at the axis for odd functions and homogeneous
Neumann boundary conditions for even functions. In addition, we have that 
the function $q$ defined by \eqref{eq:70} should vanished at the axis
\begin{equation}
  \label{eq:9b}
  q|_{\rho=0}=0.
\end{equation}
Since $q$ is an even functions, from \eqref{eq:9b} we deduce that $q=O(\rho^2)$
near the axis.  Finally, there is an important regularity condition which comes
from the axial regularity of the 3-dimensional extrinsic curvature. Let us
define the following quantity
\begin{equation}
  \label{eq:32}
 w= \frac{1}{\rho} \left(-\frac{\eta'}{\eta}+\chi_{\rho\rho}\right). 
\end{equation}
Then it follows that
\begin{equation}
  \label{eq:20}
  w=O(\rho),
\end{equation}
near the axis. This is the equivalent of the regularity condition given in
equation (50) in \cite{Rinne:2005sk} adapted to our conformally rescaled
metric. See also \cite{Rinne:thesis}, \cite{Choptuik:2003as}.

The fall off conditions at infinity are the standard asymptotically flat
ones. In particular we have
\begin{equation}
  \label{eq:21}
  \lim_{r\to \infty} \bar \alpha =1, 
\end{equation}
and
\begin{equation}
  \label{eq:9}
 \sigma,\,  \beta^A =O(r^{-1}),\quad \chi_{AB}=O(r^{-2}),
\end{equation}
as $r \to \infty$.

\section{Linearized equations}
\label{sec:linearized-equations}

In this section we make a linear expansion around Minkowski of the Einstein
equations in the maximal-isothermal gauge described in the previous
section. Note that for Minkowski we have
\begin{equation}
  \label{eq:81}
  \eta=\rho^2,
\end{equation}
and hence, due to the rescaling (\ref{eq:17}), the
background metric $h_{ab}$, given in coordinates by \eqref{eq:26}, is non-flat
\begin{equation}
  \label{eq:55}
  h=\rho^2\left(- dt^2+ d\rho^2 +dz^2 \right).
\end{equation}
The other background quantities are given by 
\begin{equation}
  \label{eq:54}
  \omega=0, \quad \alpha=\rho, \quad \beta^A=0, \quad \chi_{AB}=0,
\end{equation}
and 
\begin{equation}
  \label{eq:57}
  u=\ln \rho, \quad q=0, \quad \sigma=0.
\end{equation}
The Hamiltonian constraint and the equation for the lapse are
non-trivial for the metric (\ref{eq:55}), namely
\begin{equation}
  \label{eq:58}
  \Ld \alpha  =0, \quad  \Ld u =\frac{2}{\rho^2}.
\end{equation}

Let us proceed with the linearization. For simplicity we will consider only the
case $\omega=0$. The first step is to compute the
lapse function. The right hand side of Eq. (\ref{eq:49})  is second-order,
then, using the boundary condition \eqref{eq:21} we obtain
\begin{equation}
  \label{eq:30}
  \bar \alpha=1. 
\end{equation}
That is, the maximal condition for the lapse  is trivial at the linearized
level. On the contrary, as we will see, the equation for the shift plays a
crucial role. 

The next step is to compute the linearization of the wave Eq. (\ref{eq:137})
 for $\sigma$, we obtain
\begin{equation}
  \label{eq:73}
  - \dot p + \Ldt \sigma=0,
\end{equation}
where dot means partial derivative with respect to $t$ and we have defined
\begin{equation}
  \label{eq:75}
  p= \partial_t \sigma -\frac{2\beta^\rho}{\rho}.
\end{equation}
In order to close the system we need an equation for $\beta^\rho$. 
Using equation  (\ref{eq:65}) and  \eqref{eq:98} 
for the momentum  we obtain
\begin{equation}
  \label{eq:77}
  \partial^A  \chi_{AB}=\bar J_A
\end{equation}
with 
\begin{equation}
  \label{eq:76}
 \bar  J_A= - p  \partial_A \rho,
\end{equation}
We define the vector field $v^A$ by Eq. (\ref{eq:46})
and then by Eq. (\ref{eq:47}) we obtain
\begin{equation}
  \label{eq:93}
  \Delta v_A=-p \partial_A \rho.
\end{equation}
From (\ref{eq:76}) we deduce $J_z=0$ and hence we get
\begin{equation}
  \label{eq:114}
  \Delta v_z=0.
\end{equation}
By the fall off condition \eqref{eq:9}, we obtain
\begin{equation}
  \label{eq:115}
   v_z=0.
\end{equation}
In the following, to simplify the notation we set
\begin{equation}
  \label{eq:31}
  v\equiv v_\rho. 
\end{equation}
Eq. (\ref{eq:93}) reads 
\begin{equation}
  \label{eq:135}
   \Delta v=-p .
\end{equation}
Using (\ref{eq:46})  we also obtain
\begin{equation}
  \label{eq:56}
  \chi_{\rho\rho}=\partial_\rho v, \quad \chi_{\rho z} = \partial_z v.
\end{equation}
For the shift we have the equation
\begin{equation}
  \label{eq:69}
 (\ck \beta)^{AB} = 2  \frac{\hat \chi^{AB}}{\rho}. 
\end{equation}
Taking a divergence to this equation (or linearizing (\ref{eq:50})) we obtain
\begin{equation}
  \label{eq:94}
\Delta \beta^A = 2\partial_B\left( \frac{\hat \chi^{AB}}{\rho}\right) 
\end{equation}
Note that in (\ref{eq:94}) we get an equation for $\beta^\rho$ decoupled from
$\beta^z$. Using
Eq. (\ref{eq:76}), (\ref{eq:77}) from this equation we get 
\begin{equation}
  \label{eq:12}
  \Delta  \beta^\rho = -\frac{2}{\rho} \left( p
    +\frac{\partial_\rho v }{\rho}\right).
\end{equation}
Eq. \eqref{eq:12}, together with (\ref{eq:135}) and (\ref{eq:73}) form a
complete system for the variables $v$, $\sigma$ and $\beta^\rho$. 
Alternative, using Eq. \eqref{eq:69} and \eqref{eq:77} we can eliminate
$\chi_{AB}$ and hence also $v$.  We get the following equation for $\beta^A$
\begin{equation}
  \label{eq:24}
  \partial_B\left (\rho (\ck \beta)^{AB}\right)=-2p\partial^A \rho. 
\end{equation}
Eq. \eqref{eq:24}, together with (\ref{eq:135}) and (\ref{eq:73}) form a
complete system for the variables $\sigma$,  $\beta^\rho$ and $\beta^z$. 

There is
however an important difficulty. The linearization of the regularity condition
\eqref{eq:32}--\eqref{eq:20} is given by 
\begin{equation}
  \label{eq:91}
  w=-\frac{1}{\rho}\left(p+\frac{\partial_\rho v }{\rho} \right), \quad  w=
  O(\rho). 
\end{equation}
Were we have used Eq. (\ref{eq:56}). From the set of equations presented
above, it is difficult to ensure that this condition will be satisfied.
To enforce this condition we will write the
equations in terms of different variables.  In order to do that, we need first
to compute the remaining equations, namely the Hamiltonian constraint and the
evolution equation for the metric and second fundamental form.
Since $\epsilon$ defined in \eqref{eq:102} is second-order, 
the Hamiltonian constraint \eqref{eq:101}  is given by
\begin{equation}
  \label{eq:74}
  \Ld q= - \Ldt\sigma.
\end{equation}
The evolution equation for $q$ is obtained from (\ref{eq:164})
\begin{equation}
  \label{eq:80}
   \dot q + \dot \sigma=\frac{1}{2}\partial_A \beta^A+\frac{\bar \beta^\rho}{\rho}.
\end{equation}
The evolution equation for $\chi_{AB}$ is obtained linearizing (\ref{eq:140})
\begin{equation}
  \label{eq:79}
  \dot \chi_{AB}= 2 \partial_{(A}q \partial_{B)}\rho-\delta_{AB} \partial_\rho
  q.
\end{equation}
We can also write the evolution Eqs. (\ref{eq:79})  in components
\begin{equation}
    \label{eq:78}
    \dot \chi_{\rho\rho}= \partial_\rho q, \quad  \dot \chi_{\rho z}= \partial_z q.
  \end{equation}
Using Eqs. (\ref{eq:56}) we deduce the important relation
\begin{equation}
  \label{eq:117}
  \dot v  = q,
\end{equation}
which only holds in the twist free case. This relation simplify the equations
considerable. From  Eq. (\ref{eq:79}) we also deduce
\begin{equation}
  \label{eq:44}
  \partial^B \dot \chi_{AB}=\Ld q\partial_A\rho.
\end{equation}

With these equations we can compute the time derivative of $w$
\begin{equation}
  \label{eq:104}
   \dot w = \frac{1}{\rho}\left(\Ld q - \frac{\partial_\rho q}{\rho}
  \right), 
\end{equation}
and hence  the evolution equation for $q$ is given by
\begin{equation}
  \label{eq:105}
  \dot q = \rho w + \rho \partial_\rho \left(\frac{\beta^\rho}{\rho} \right).
\end{equation}
As a consequence, $q$ satisfies the following wave equation
\begin{equation}
  \label{eq:118}
 \ddot q =\Ld q-\frac{\partial_\rho q}{\rho}+ \rho \partial_\rho \left(
  \frac{\dot \beta^\rho}{\rho} \right ). 
\end{equation}
We also have
\begin{equation}
  \label{eq:92}
  \Ld \dot \beta^\rho= \frac{2}{\rho}\left(\Ld q - \frac{\partial_\rho q}{\rho}
  \right).
\end{equation}
Eqs. \eqref{eq:118} and \eqref{eq:92} form a complete systems for the
variables $q$ and $\beta^\rho$.  A similar choice of variable was used in by
\cite{Choptuik:2003as} \cite{Rinne:2008tk}\cite{Garfinkle:2000hd}. However, in
our particular case (i.e. linear equation without twist) it is possible a
further simplification, namely to use Eq. \eqref{eq:117} and hence replace
$q$ by $v$ in these equations and then integrate in time.  In this way we
obtain our main Eqs. \eqref{eq:119b} and \eqref{eq:120b}. The advantage of
using $v$ as a variable is that the mass integral has a simple expression in
terms of $v$ given by \eqref{eq:124}. This formula for the mass is obtained
expanding up to second-order the energy density \eqref{eq:102}, using equations
\eqref{eq:56} to replace $\chi_{AB}$ and Eq. \eqref{eq:135} to replace
$\eta'/\eta$. We discuss this in more detail in the next section. 

The boundary conditions \eqref{eq:84} and \eqref{eq:146} for $\beta^\rho$ arise
from the axial regularity condition \eqref{eq:9c} and the asymptotically flat
fall off \eqref{eq:9}. Conditions \eqref{eq:148} arise from the axial
regularity conditions for $q$ given by \eqref{eq:10c} and \eqref{eq:9b}.  The
main advantage of Eqs. \eqref{eq:119b} and \eqref{eq:120b} is that they have
build in the regularity condition \eqref{eq:91} as we will see in the next
section. Let us mention that the non-trivial regularity condition \eqref{eq:91}
is written in terms of $v$ as follows
\begin{equation}
  \label{eq:22}
   w=\frac{1}{\rho}\left(\Ld v -\frac{\partial_\rho v }{\rho} \right), \quad  w=
  O(\rho). 
\end{equation}

The component $\beta^z$, which does not appears in
Eqs. \eqref{eq:119b} and \eqref{eq:120b}, can be calculated using 
\begin{align}
  \label{eq:122}
  \partial_\rho \beta^\rho - \partial_z \beta^z & = 2\frac{\partial_\rho v
  }{\rho},\\
\partial_z \beta^\rho + \partial_\rho \beta^z & = 2\frac{\partial_z v
  }{\rho}.\label{eq:122b}
\end{align}
Or, alternative, using Eq. \eqref{eq:110} which is is obtained taking a
derivative to Eqs. \eqref{eq:122}--\eqref{eq:122b}.
Finally, the four dimensional perturbation \eqref{eq:106} is obtained using the
line element \eqref{eq:128} and the background values \eqref{eq:81},
\eqref{eq:54}, \eqref{eq:57}.

\section{Properties of the linear equations}
\label{sec:prop-line-equat}
In this section we analyze some properties of our linear equations
\eqref{eq:119b} and \eqref{eq:120b}.  We begin with the symmetries of these
equations.  The first symmetry is given by translation in $z$. This is to be
expected since the gauge fixes the axis (and hence there is no translation
freedom in $\rho$), but we still have the freedom to chose the origin in the
$z$ coordinate. Then, if we have a solution $v, \beta^\rho$; the derivative
$\partial_z v, \partial_z \beta^\rho$ is also a solution, since $\partial_z$
commute with all the differential operators because their coefficients depend
only on $\rho$.  The same argument applies to time translations, which is the
second symmetry of the equations.  The third symmetry
is scaling. Let $s$ a positive real number. For a given solution $v(t,\rho,z)$
we define the rescaled function as
\begin{equation}
  \label{eq:173}
  v_s(\hat t, \hat \rho, \hat z)=v\left(\frac{t}{s},\frac{\rho}{s},\frac{z}{s} \right), 
\end{equation}
where
\begin{equation}
  \label{eq:174}
  \hat t =\frac{t}{s} , \quad \hat \rho=  \frac{\rho}{s}, \quad \hat z  = \frac{z}{s}.
\end{equation}
And the same for $\beta^\rho$. Then, $v_s$ define also a solution in terms of
the rescaled coordinates. The mass rescales like
\begin{equation}
  \label{eq:175}
  m\to s m.
\end{equation}

In order to understand the equations in a simpler situation, let us first
consider Eqs. \eqref{eq:119b} and \eqref{eq:120b} on a bounded domain
$\Omega$ which does not contain the axis. On $\Omega$ the coefficient of
Eqs. \eqref{eq:119b} and \eqref{eq:120b} are smooth. Eq. \eqref{eq:120b}
is an elliptic equation for $\beta^\rho$ if we consider $v$ as a given
function.  Hence, in order to solve this equation we need to prescribe elliptic
boundary conditions for $\beta^\rho$ on $\partial \Omega$. For example,
Dirichlet or Neumann boundary conditions. Eq. \eqref{eq:119b} is a wave
equation for $v$ if we consider $\beta^\rho$ as a given function. To solve this
wave equation we need to prescribe initial data for $v$ and $\dot
v$ at $t=0$ together with compatible boundary conditions for $v$ at $\partial
\Omega.$ For example Dirichlet, Neumann or Sommerfeld
boundary conditions for $v$ at $\partial \Omega$.  The equations are of course
coupled, so it is not obvious that the above procedure of fixing boundary
conditions  is  correct since $v$ and $\beta^\rho$ are not ``given
functions''. However, it is possible to prove that this is procedure is in fact
correct. Consider the following iteration scheme
\begin{align}
  \label{eq:19}
  \ddot v_{n+1}  -\Ld v_{n+1}  + \frac{\partial_\rho v_{n+1} }{\rho}  &= 
  \rho \partial_\rho \left(\frac{\beta_n^\rho}{\rho}  \right),\\
  \label{eq:19b}
  \Ld \beta_{n+1}^\rho & = \frac{2}{\rho} \left(\Ld v_n - \frac{\partial_\rho
      v_n}{\rho}\right).  
\end{align}
In this iteration, the equations are not coupled and hence the boundary
conditions mentioned above (which are kept fixed) are correct.  Following
similar arguments to the one presented in \cite{ChoquetBruhat:1996ik} (see also
\cite{kreiss89}) it is not difficult to see that this iteration converges for
some small time interval.  And hence we get well-posedness for the linear system
\eqref{eq:119b} and \eqref{eq:120b} under these boundary conditions on the
domain $\Omega$. The reason why the iteration \eqref{eq:19} and \eqref{eq:19b}
converges is the following. From Eq. \eqref{eq:19b}, using standard elliptic,
estimates we obtain that $\beta^\rho$ is equivalent (in number of derivatives)
to $v$. Hence, the term containing $\beta^\rho$ in Eq. \eqref{eq:19} is
equivalent to a first order derivative of $v$ and then it is not in the
principal part of the wave equation. This rough argument suggests that the
combination of elliptic estimates and energy estimates for the wave equations
will close and hence the iteration will converge.  This is basically the
argument presented in \cite{ChoquetBruhat:1996ik} and \cite{kreiss89} . If the
domain $\Omega$ is not bounded, this argument will still work if we add
appropriate fall-off conditions at infinity. However, the situation change
drastically when $\Omega$ includes the axis. Let us analyze that case.

Since the axis is a singular boundary for the equations, we are not free to
chose arbitrary boundary condition there. In fact $\beta^\rho$ and
$\partial_\rho v$ should vanishes at the axis, otherwise the equations become
singular. If we use L'H\^opital rule, we conclude that the term with $\beta^\rho$
in \eqref{eq:119b} contain in fact two derivatives with respect to $\rho$ at
the axis. That is, due to the L'Hopital limit, to divide by $\rho$ is
equivalent as to take a derivative with respect to $\rho$ at the axis.  But
then, using Eq. \eqref{eq:120b}, we conclude that this term is equivalent
to second derivatives of $v$ and hence it is in the principal part of the wave
equation.  We can not conclude that the iteration scheme
\eqref{eq:19}--\eqref{eq:19b} converges if we include the axis in the
domain. This is, roughly speaking, the main difficulty to prove the
well-posedness of
the linear system \eqref{eq:119b} and \eqref{eq:120b}.  It appears to be difficult
to identify the principal part of the system at the axis and to construct an
appropriate iteration scheme.

Let us discuss in detail the boundary conditions at the axis. We are interested
in solutions $v$ which vanish at the axis, this comes from the regularity
condition \eqref{eq:9b}. Moreover, we have seen in section
\ref{sec:bound-cond-axial}, the smoothness of the spacetime metric at the axis
implies that the functions $\beta^\rho$ and $v$ should satisfies the parity
conditions \eqref{eq:10c} and \eqref{eq:9c}. Let us see heuristically, how
these conditions are automatically implied by the equations provided we impose
the following standard boundary conditions.  At the axis we impose
\begin{equation}
  \label{eq:39}
  \beta^\rho|_{\rho=0}=0. 
\end{equation}
For $v$ we prescribe initial data 
\begin{equation}
  \label{eq:82}
  v|_{t=0}=f, \quad \dot v|_{t=0}=g,
\end{equation}
such that
\begin{equation}
  \label{eq:63}
    f|_{\rho=0}=0, \quad  g|_{\rho=0}=0.
\end{equation}
Note that we are not imposing any condition on $v$ at the axis for $t>0$.  We
make a formal series expansion, namely let as assume that our solution is
smooth at the axis and has the form
\begin{equation}
  \label{eq:27}
  v=\sum_{n=0}^\infty \rho^n a_n(t,z), \quad  \beta^\rho=\sum_{n=0}^\infty \rho^n b_n(t,z). 
\end{equation}
Substituting these expansions in Eqs. \eqref{eq:119b} and \eqref{eq:120b} we
obtain the following recurrence relation for the coefficients
\begin{equation}
  \label{eq:33}
  \ddot a_n =(n+2)n a_{n+2}+\partial^2_z a_n+n b_{n+1},
\end{equation}
and
\begin{equation}
 \label{eq:33b}
(n+1)nb_{n+1}+ \partial^2_z b_{n-1} =2(n+2)na_{n+2} + \partial^2_z a_n.
\end{equation}
These expressions are valid for all integer $n$, with the convention that the
coefficients $b_n$ and $a_n$ vanished for $n<0$. 
The first non-trivial $n$ in Eq. \eqref{eq:33} is $n=-1$, which gives the relation
\begin{equation}
  \label{eq:34}
  a_1+b_0=0. 
\end{equation}
The term  $n=0$ is given by
\begin{equation}
  \label{eq:35}
   \ddot a_0 =\partial^2_z a_0.
\end{equation}
This is a wave equation in 1-dimension. 
From  the boundary conditions \eqref{eq:39} we obtain
\begin{equation}
  \label{eq:61}
  b_0=0.
\end{equation}
Hence we deduce from \eqref{eq:34} that 
\begin{equation}
  \label{eq:116}
  a_1=0.
\end{equation}
From the initial data conditions
\eqref{eq:63} we have that
\begin{equation}
  \label{eq:108}
  a_0|_{t=0}=0 , \quad  \dot a_0|_{t=0}=0.  
\end{equation}
These provides trivial initial data for the wave Eq. \eqref{eq:35} and
hence we deduce
\begin{equation}
  \label{eq:113}
  a_0=0.
\end{equation}
That is, we have deduced the behavior $v=O(\rho^2)$ only from the boundary
conditions \eqref{eq:39} and the condition on the initial data \eqref{eq:82}. 
We want to prove now that \eqref{eq:61} and \eqref{eq:116} imply that all $a_n$
with $n$ odd and all $b_n$ with $n$ even are zero. We prove this by
induction. Let us assume that for some $n$ (with $n\geq 1$) we have that
\begin{equation}
  \label{eq:139}
  b_{n-1}=0, \quad a_n=0. 
\end{equation}
Using Eq. \eqref{eq:33b} we deduce
\begin{equation}
  \label{eq:143}
  (n+1)b_{n+1}=2(n+2)a_{n+2},
\end{equation}
and from \eqref{eq:33} we have
\begin{equation}
  \label{eq:144}
   (n+2)a_{n+2}=-b_{n+1}. 
\end{equation}
And then we have $a_{n+2}=b_{n+1}=0$. Since \eqref{eq:139} is valid for $n=1$
we have proved the desired result. That is, the solutions $v$ and $\beta^\rho$
satisfy the parity conditions \eqref{eq:10c} and \eqref{eq:9c} respectively.
Using that $v$ is an even function of $\rho$ and that $v=O(\rho^2)$ it is
straightforward to deduce that the regularity condition \eqref{eq:22} holds for
all times.

We analyze the fall off behavior of the solution $v$. This behavior is
completely determined by the initial data $f$ and $g$. Let us assume that the
initial data has compact support. In the case of the wave equation, the signal
will propagate with finite speed and hence the solution will always have compact
support for any finite time. In our case, however, the coupling with the
elliptic Eq. \eqref{eq:120b} produce a non-local behavior. Even if we
start with compactly supported data, the function $\beta^\rho$ will
instantaneously spread to all space.  Let us perform a formal expansion in $r$
to see the typical behavior of $v$. We have that $\beta^\rho=O(r^{-1})$ for
all times, this is prescribed by the boundary conditions. In \cite{Dain:2008xr}
it has been proved that this implies that $\beta^\rho/\rho= O(r^{-2})$, and
hence the terms containing $\beta^\rho$ in \eqref{eq:119b} is
$O(r^{-2})$. Then, at $t=0$ we obtain that $\ddot v=O(r^{-2})$. If we take time
derivatives of the equations and repeat this argument, we get that all time
derivatives of $v$ are $O(r^{-2})$. Then, we conclude that the typical fall-off
behavior for asymptotically flat solutions is given by \eqref{eq:29}, in the
sense that we can not expect a faster decay in general.  Instead of
compactly supported data we can begin with initial data for $v$ such that they
are $O(r^{-2})$ at infinity. 

Let us discuss now the most important property of equations
\eqref{eq:119b} and \eqref{eq:120b} namely the mass conservation. As usual, the
mass appears as a second-order quantity that can be calculated in terms of
squares of first order quantities.  The density (\ref{eq:102}) up to this order
is given
\begin{equation}
  \label{eq:100}
  \epsilon= 4\frac{|\partial v|^2}{\rho^2}+(\Ld v)^2 +
    |\partial \sigma |^2.
\end{equation}
The total mass is calculated by the integral (\ref{eq:103}).  The mass integral
is conserved for the full nonlinear equations in this gauge (see
\cite{Dain:2008xr}) and hence it is conserved at the linearized level. It is
however important to compute explicitly this conservation formula using only
the linear Eqs. (\ref{eq:119b}) and (\ref{eq:120b}).  Note that the
function $\sigma$ appears in the mass and this function should be calculated
from $v$ using Eq. \eqref{eq:86}.  To compute the time derivative of $m$
we need first to calculate the time derivative of $\sigma$.  Using the
evolution Eq. (\ref{eq:119b}) only, we compute
\begin{equation}
  \label{eq:187}
  \Ldt (-\Ld v + 2\frac{\beta^\rho}{\rho})= -\Ld \ddot v
  +\frac{1}{\rho} \partial_\rho \left[ \rho\left(\Ld \beta^\rho -2 L(v)\right) \right],
\end{equation}
where we have defined
\begin{equation}
  \label{eq:83}
  L(v)=\partial^A\left( \frac{\partial_A v}{\rho}\right)= 
\frac{1}{\rho} \left(\Ld v - \frac{\partial_\rho v}{\rho}\right). 
\end{equation}
Then, using \eqref{eq:120b} and   the time derivative of  \eqref{eq:86} we get
\begin{equation}
  \label{eq:189}
   \Ldt (-\Ld v + 2\frac{\beta^\rho}{\rho} -\dot \sigma)=0.
\end{equation}
If we are solving in the whole half plane $\Rdm$
then, by the fall-off conditions,  we deduce that the only possible
solution of this equation is the trivial one, and hence
\begin{equation}
  \label{eq:190}
  \dot \sigma= -\Ld v + 2\frac{\beta^\rho}{\rho}.
\end{equation}

We have proved that Eqs. (\ref{eq:119b}) and (\ref{eq:120b}) together with
(\ref{eq:86}) imply Eq. (\ref{eq:190}). 
We can also formulate the system in a different way.  We can take
(\ref{eq:119b}) and (\ref{eq:120b}) and Eq. (\ref{eq:190}), instead
of (\ref{eq:86}),   as an evolution
equation for $\sigma$.  If we take the Laplacian $\Ldt$ to both
sides of Eq. (\ref{eq:190}) and use the identity \eqref{eq:187} together
with Eq. (\ref{eq:120b})  we obtain
\begin{equation}
  \label{eq:23}
  \Ldt \dot \sigma=-\Ld \ddot v.
\end{equation}
Hence, if we chose  initial condition for $\sigma$ such that
\begin{equation}
  \label{eq:62}
 \Ldt \sigma|_{t=0} = -\Ld \dot v|_{t=0},  
\end{equation}
Eq. \eqref{eq:23} implies (\ref{eq:86}).  This two different ways of
calculating $\sigma$ correspond to a constrained system and a free system
(using the terminology defined in \cite{Rinne:thesis}). The previous
calculation is nothing but the propagation of the Hamiltonian constraint at the
linearized level. For the full Einstein equations, the difference of
constrained and free evolution schemes involves different set of evolutions
equations. In our linear system the evolution equations are the same (namely,
(\ref{eq:119b}) and (\ref{eq:120b})), the difference is the way the function
$\sigma$ (and hence the mass) is calculated. These two ways are of course
completely equivalent when the domain is the whole half plane $\Rdm$, however,
as we will see, they are not equivalent for a bounded domain.

Using Eqs. (\ref{eq:190}), (\ref{eq:86}), (\ref{eq:119b}) and (\ref{eq:120b}) we
obtain the following local conservation law for the density $\epsilon$ defined
by \eqref{eq:100}
\begin{equation}
  \label{eq:89}
 \rho \dot \epsilon =\partial_A \epsilon^A,
\end{equation}
where
\begin{equation}
  \label{eq:99}
  \epsilon_A= 8 \frac{\partial_A v}{\rho}\dot v+
  2 \rho\dot \sigma \partial_A \sigma + 4 \beta^\rho \partial_A\dot v
  -4\dot v \partial_A \beta^\rho. 
\end{equation}
The vector $\epsilon^A$ can be interpreted as the energy flow of the
gravitational field. 
If we integrate Eq. (\ref{eq:89}) in $\Rdm$ we have that the
boundary terms vanishes both at the axis (by the axial regularity) and at
infinity (by the fall off conditions). Then we have
\begin{equation}
  \label{eq:41}
  \dot m =0. 
\end{equation}
We can also integrate Eq. (\ref{eq:89}) on a bounded domain $\Omega$,
namely we define the mass contained in $\Omega$ by 
\begin{equation}\label{eq:mass_omega}
m_\Omega=\int_\Omega \epsilon \rho \, d\rho dz, 
\end{equation}
and then we have
\begin{equation}
  \label{eq:53}
  \dot m_\Omega= \oint_{\partial \Omega} \epsilon^An_A,
\end{equation}
where $n^A$ is the unit normal of $\partial \Omega$. The quantity
$\epsilon^An_A$ measure how much energy is leaving or entering the domain.  The
local conservation formula (\ref{eq:89}) can be generalized for the non-linear
equations \cite{Dain09}.

Using the conservation of the mass (\ref{eq:41}) we can prove uniqueness of
solutions of the system.  Let us say we have two different solutions with the same
initial data. We take the difference between the two solutions.  The difference
satisfies the same equation with zero initial data. In particular $\sigma$ on
the initial surface is zero. And hence the mass is zero. Since it is conserved
the mass is zero for all times, which implies that the solution is zero.

In the case of hyperbolic equations (the wave equation for example) the
conservation of the energy gives also local properties of the solution, namely
finite speed propagation of signals.  However this is not the case here; the
elliptic equation implies a non-local behavior of the solution.

\begin{figure}
\begin{center}
\includegraphics[width=4cm]{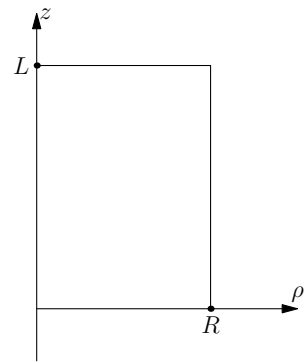}
\caption{The bounded domain $\Omega$ for the numerical evolution \label{fig:1}}
\end{center}
\end{figure}

The discussion above applies for the domain $\Rdm$ which is the
relevant domain for the equations. However, in numerical computation we need to
solve the equations on a finite grid and hence it is necessary to impose
boundary conditions on a bounded domain. A typical domain for the numerics is
shown in Fig. \ref{fig:1}. As we mention in section \ref{sec:main-results},
for our present purpose we only need to prescribe some boundary conditions
compatible with asymptotic flatness. For example, homogeneous Dirichlet boundary
conditions for $v$ and $\beta^\rho$. 
However, the mass formula rise an interesting point here. On a bounded domain,
to calculate $\sigma$ we have two possibilities.  First, we can
 determine $\sigma$ as the unique solution of the elliptic equation
 (\ref{eq:86})   with some
 boundary conditions.
If we do so, then we again deduce 
Eq. (\ref{eq:189}). However, from this equation we can not deduce
(\ref{eq:190}). In effect, we have
\begin{equation}
  \label{eq:48}
  \dot \sigma + \Ld v - 2\frac{\beta^\rho}{\rho}=H,
\end{equation}
where $H$ satisfies
\begin{equation}
  \label{eq:51}
 \Ldt H =0.  
\end{equation}
We can not conclude that $H$ is zero from this equation, because $H$
will have non-trivial boundary condition. Namely, let us assume the we
prescribe some boundary condition for $\sigma$. We can not control the boundary
value of $\Ld v$, and hence we can not ensure that $H$ vanishes at the
boundary. 
In fact, the function $H$ is fixed as the unique solution of \eqref{eq:51} with
boundary values
\begin{equation}
  \label{eq:59}
  H |_{\partial \Omega}= ( \dot \sigma + \Ld v -
  2\frac{\beta^\rho}{\rho})|_{\partial \Omega}. 
\end{equation}
Then, if we compute the time derivative of the density $\epsilon$ we get
\begin{equation}
  \label{eq:60}
 \rho   \dot \epsilon =\partial_A\epsilon^A +  \rho H \Ld \dot v.
\end{equation}
That is, we do not get a conservation law, there is a volume term given by
$H$.  There seems
to be no boundary conditions for $\sigma$ that can ensure $H$ to vanishes. 

The other possibility is to compute $\sigma$ using the evolution equation
(\ref{eq:190}) 
with initial condition (\ref{eq:62}). From (\ref{eq:190}), in the same way as
we mentioned above we deduce \eqref{eq:23}, since in this deduction the
boundary conditions play no role. Using the initial data condition
(\ref{eq:62}), from   \eqref{eq:23} we deduce \eqref{eq:86}. 
That is, we are in the same situation as the whole domain. Hence, in this case
we recover (\ref{eq:89}), where $\epsilon^A$ is given by the same expression
\eqref{eq:99}. From this point of view, this evolution scheme appears to be
better than the previous one. 

In this scheme, we are free to chose any elliptic boundary condition for
$\beta^\rho$ and any boundary condition for $v$ compatible with the wave
equation. For $\sigma$ we do not have any freedom, and hence we can not
prescribe the boundary value of this function.

A natural choice of boundary conditions would be to force the boundary integral
in \eqref{eq:53} to have a definite sign. These conditions would have the
interpretation of radiative boundary conditions, in the sense that the energy is
leaving the domain. To prescribe such conditions seems not to be possible (at
least for generic data) since
we do not have any control on the term with $\sigma$. However, we can do
something intermediate. Namely, if we impose
Sommerfeld boundary condition for $v$
\begin{equation}
  \label{eq:90}
 \dot v= -n^A\partial_Av,
\end{equation}
and homogeneous Dirichlet conditions for $\beta^\rho$ we have that the first
term in \eqref{eq:99} has negative sign, the third term is zero. For the second
and fourth term we have no control a priori. 
But we can expect that the influence of these term is small at least for some
class of initial data. If this is true, then we get 
\begin{equation}
  \label{eq:111}
  \dot m_\Omega \leq 0. 
\end{equation}
This is what we observe in our numerical simulations described in the next
sections.

\section{Numerical implementation}
\label{sec:numerical-techniques}
In this section we want to study numerically the initial-boundary value problem
(IBVP) for the Eqs. (\ref{eq:119b}) and (\ref{eq:120b}). In this problem the
symmetry axis, $\rho=0,$ becomes a boundary of our domain. Notice then, that 
working with the variable $v$ poses an inconvenient as regards the boundary
condition at $\rho=0$ since, according to \eqref{eq:148}, this
function satisfies both, homogeneous Dirichlet boundary condition and
homogeneous Neumann boundary condition. It is then convenient to rewrite the
equations in terms of a new variable for which the smoothness properties at the
symmetry axis defines a unique, equivalent, boundary condition. We define
$\bv=v/\rho.$ This new variable vanishes linearly with $\rho$ and the correct
boundary condition is simply homogeneous Dirichlet at $\rho=0.$
 
The equation for $\bv(\rho,z,t),$ with $\rho\in[0,R],~z\in[0,L],$ and $t\ge 0,$ is 
\begin{equation}\label{eq:hyp}
\ddot \bv=\Delta \bv + \partial_\rho \Bigl(\frac{\bv}{\rho}\Bigr) +
\partial_\rho\Bigl(\frac{\bbeta}{\rho}\Bigr),
\end{equation}
where $\bbeta(\rho,z,t)$ is determined by the elliptic equation
\begin{equation}\label{eq:ell}
\Delta\bbeta=2\left(\Delta \bv + \partial_\rho
\Bigl(\frac{\bv}{\rho}\Bigr)\right)
\end{equation}
 with homogeneous Dirichlet boundary conditions, 
\begin{equation}\label{bc:beta}\begin{split}
\bbeta(0,z,t)&=\bbeta(R,z,t)=0,\quad z\in[0,L],\\
\bbeta(\rho,0,t)&=\bbeta(\rho,L,t)=0,\quad \rho\in[0,R].
\end{split}
\end{equation}
for all $t\in[0,\infty).$

The boundary condition for $\bv$ at the symmetry axis is
\begin{equation}\label{bc:axis}
\bv(0,z,t)=0,
\end{equation}
while at the outer boundaries we study two possibilities, homogeneous
Dirichlet,
\begin{equation}\label{bc:dir}
\bv(R,z,t)=\bv(\rho,0,t)=\bv(\rho,L,t)=0,
\end{equation}
or Sommerfeld (outgoing waves)
\begin{equation}\label{bc:som}\begin{split}
\dot\bv(R,z,t)&= -\partial_\rho\bv(R,z,t),\\
\dot\bv(\rho,0,t)&= \bv_z(\rho,0,t),\\
\dot\bv(\rho,L,t)&= -\bv_z(\rho,L,t).
\end{split}
\end{equation}
The initial data are
\begin{equation}\label{idata}\begin{split}
\bv(\rho,z,0)&=\bv_0(\rho,z),\\
\dot\bv(\rho,z,0)&=\bv_{0t}(\rho,z),
\end{split}
\end{equation}
where $\bv_0$ and $\bv_{0t}$ are $C^\infty$ functions with compact support in
$(0,R)\times(0,L)$ so that the compatibility of the boundary and initial data is
not an issue.

The Eqs. (\ref{eq:hyp})--(\ref{idata}) constitute the IBVP we approximate
with our finite difference scheme.

We want to emphasize here an important difference between our numerical approach with
the usual approaches in the area (see for example
\cite{Choptuik:2003as}). We solve the IBVP for (\ref{eq:hyp}) as a second-order
equation just it is written above, i.e. we do not reduce 
(\ref{eq:hyp}) to a first order system of equations. The treatment of evolution
equations as second-order equations as opposed to first order systems of
equations has several advantages.  For example, the number of dynamical fields,
and then the number of equations, is not increased. This facilitate the
treatment of the boundary conditions.  There are also numerical accuracy
advantages. In the context of general relativity, this has been stressed in
\cite{Kreiss:2004}. In particular the simplest proofs of well-posedness for
general initial-boundary value problems for Einstein's equations have been found
recently using second-order systems of equations \cite{Kreiss:2006mi},\cite{Kreiss:2007zz}.

\paragraph{The Implementation.} In our numerical experiments we always consider
square domains, i.e., $R=L.$ To define the numerical grid let $N$ be a positive
integer and $h=L/N$ the space stepsize. We define our grid to be half a stepsize
displaced from all the boundaries. We think of our grid as a uniformly
distributed set of points each of which is at the center of one of the $N^2$
square cells covering the domain. The coordinates of the gridpoint at the site
$(i,k)$ are then
\begin{align}
\rho_i&= h(i-3/2), \quad i=0,1,2,\dots N+3\\
z_k&= h(k-3/2), \quad k=0,1,2,\dots N+3
\end{align}
The sites $(i,k)$ with $2\le i,k\le N+1$ are within the domain, while the sites
with $i=0,1,N+2,N+3$ and $k=0,1,N+2,N+3$ are ``ghost points'' used to ease the
implementation of the boundary conditions \cite{Rinne:thesis}. Time is
discretized as
\begin{equation}\label{time}
t_n= n\delta t, \qquad n=0,1,2,3,\dots
\end{equation}

We use capital latin letters to denote the grid functions associated to the
dynamical variables. Also, we use sub-indices to denote the space-site indices
and a super-index to denote the time step. This is,
\begin{equation}\label{gridfunctions}\begin{split}
V^n_{i,k}\quad \mbox{corresponds to}~\bv(\rho_i,z_k,t_n),\\
B^n_{i,k}\quad \mbox{corresponds to}~\bbeta(\rho_i,z_k,t_n),
\end{split}
\end{equation}

Besides the uniform grid we introduce the extra gridpoints placed at the
physical boundary
\[
(\rho=L,z_k),\quad (\rho_i,z=0),\quad (\rho_i,z=L)
\]
and denote the values of $\bv$ at these points as
\begin{equation}\label{boundary_values}
\bV^n_{L,i},\quad \bV^n_{i,0}, \quad \bV^n_{i,L}
\end{equation}
respectively.

In our difference scheme we approximate space derivatives by the standard fourth
order accurate centered difference operators given by \cite{Gustafsson95}
\begin{equation}\label{4thdiff}\begin{split}
D &:=D_0\Bigl(I-\frac{h}{6}D_+D_-\Bigr)\\
D^2 &:=D_+D_-\Bigl(I-\frac{h^2}{12}D_+D_-\Bigr)
\end{split}
\end{equation}
and add a sub-index $\rho$ or $z$ to indicate what coordinate the operator is
acting on. For example $\partial^2_z \bv(\rho_i, z_k, t_n)$ is approximated by

\begin{multline*}
D^2_z V^n_{i,k} =\\
 \frac{-V^n_{i,k-2} + 16 V^n_{i,k-1} - 30 V^n_{i,k} + 16 V^n_{i,k+1} - V^n_{i,k+2}}{12 h^2}.
\end{multline*}

At every time step we need to solve the elliptic Eq. (\ref{eq:ell})
which we approximate by
\begin{multline}\label{eq:discrete-ell}
(D^2_\rho + D^2_z)B^n_{i,k}= 2\left((D^2_\rho + D^2_z)V^n_{i,k} + D_\rho
\Bigl(\frac{V^n_{i,k}}{\rho_i}\Bigr)\right),\\
\qquad i,k=2,3,\dots N+1.
\end{multline}
We solve this difference equation iteratively using the Gauss-Seidel iteration
scheme, and stop the iteration when the difference between both sides in
(\ref{eq:discrete-ell}) is smaller, in maximum norm, than a given small
tolerance $\varepsilon.$ We then extend the solution to the ghost points---so
that the homogeneous boundary condition is satisfied---as follows
\begin{align*}
B^n_{0,k}&=-B^n_{3,k},&    B^n_{1,k}&=-B^n_{2,k}\\
B^n_{N+2,k}&=-B^n_{N+1,k},&B^n_{N+3,k}&=-B^n_{N,k}\\
B^n_{i,0}&=-B^n_{i,3},&    B^n_{i,1}&=-B^n_{i,2}\\
B^n_{i,N+2}&=-B^n_{i,N+1},&B^n_{i,N+3}&=-B^n_{i,N}
\end{align*}

We now describe how we approximate (\ref{eq:hyp}) using fourth-order accurate
difference approximations in space; to use second-order accurate approximations
instead, we just need to change $D$ and $D^2$ in what follows by $D_0$ and
$D_+D_-$ respectively.

\medskip
\underline{$t=0.$} We set 
\begin{equation}\label{step-zero}
V^0_{i,k}=\bv_0(\rho_i,z_k),\qquad i,k=2,3,\dots N+1\\
\end{equation}
and extend the solution to vanish at all ghost points and boundary points since the
initial data has compact support. Then we compute $B^0_{i,k}$ by solving
(\ref{eq:discrete-ell}) as explained above.

\medskip
\underline{$t=\delta t$} (first step). We do, for $i,k=2,3,\dots N+1,$
\begin{multline}
V^1_{i,k}=V^0_{i,k} + \delta t~\bv_{0t}(\rho_i, z_k) + \frac{1}{2}(\delta t)^2
\left((D^2_\rho + D^2_z)V^0_{i,k}\right.\\
\left.+ D_\rho(V^0_{i,k}/\rho_i) + D_\rho(B^0_{i,k}/\rho_i)\right)
\end{multline}
Now, if working with boundary condition (\ref{bc:axis}),(\ref{bc:dir}), 
we define the solution at the boundary points to vanish
\begin{equation}\label{boundary-points-dir}
\bV^1_{L,k}=\bV^1_{i,0}=\bV^1_{i,L}=0,
\end{equation}
while if working with boundary condition (\ref{bc:axis}),(\ref{bc:som}) we
evolve the boundary points by integrating the boundary condition using explicit
Euler scheme. For example for the boundary $\rho=L$
\begin{equation}
\bV^1_{L,k} = \bV^0_{L,k} -\delta t\, \tilde D^\rho \bV^0_{L,k},
\end{equation}
where
\[
\tilde D^\rho \bV^0_{L,k} = \frac{27(V^0_{N+2,k} - V^0_{N+1,k}) - (V^0_{N+3,k} -
V^0_{N,k})}{24 h}
\]
is a fourth-order accurate approximation of the normal first derivative at the border
$\rho=L.$ We now extend the solution to the ghost points as 
\begin{align*}
V^1_{0,k}&=-V^1_{3,k},& V^1_{1,k}&=-V^1_{2,k}\\
V^1_{N+2,k}&=2\bV^1_{L,k} - V^1_{N+1,k},& V^1_{N+3,k}&=2 \bV^1_{L,k}-V^1_{N,k}\\
V^1_{i,0}&=2\bV^1_{i,0} - V^1_{i,3},& V^1_{i,1}&=2 \bV^1_{i,0} - V^1_{i,2}\\
V^1_{i,N+2}&=2\bV^1_{i,L} - V^1_{i,N+1},& V^1_{i,N+3}&=2 \bV^1_{i,L} - V^1_{i,N}
\end{align*}
Finally, we compute $B^1_{i,k}$ as explained above.

\medskip
\underline{At $t=n\,\delta t.$} With $n=2,3,\dots$ we evolve the solution with
the two step method 
\begin{multline}\label{nth-step}
V^{n}_{i,k}=2 V^{n-1}_{i,k} - V^{n-2}_{i,k} + (\delta t)^2 \left((D^2_\rho +
D^2_z)V^{n-1}_{i,k} \right.\\
\left. + D^\rho(V^{n-1}_{i,k}/\rho_i) + D^\rho(B^{n-1}_{i,k}/\rho_i)\right)
\end{multline}
for $i,k=2,3,\dots N+1.$ Then we impose the boundary conditions exactly as done
in the first step. Finally we compute $B^n_{ik}$ as explained above.

We notice that the second derivative in time is approximated by $D_+D_-$ which
is second-order accurate. The time step we use in all our runs is $\delta
t=h/10$.  The ratio $\delta t/h=0.1$ satisfies the Courant condition and we see
from our runs that the whole method turns out to be numerically stable.

Besides the solution $V^n_{i,k}$ and $B^n_{i,k},$ an essential quantity we want
to compute is the mass $m_\Omega(t),$ defined by (\ref{eq:100}) and
(\ref{eq:mass_omega}), during the whole evolution. To this end we need to
compute $\sigma(t)$ on the physical domain at all times. Given the
approximations of $\bv$ and $\beta,$ we compute $\sigma(t)$ by integrating
(\ref{eq:190})---rewritten in terms of $\bv,$ as an ODE at each
gridpoint. The initial data for these ODEs is computed by solving the elliptic
Eq. (\ref{eq:62}), also rewritten in terms of $\bv,$ only once at initial
time with homogeneous Dirichlet boundary conditions and using the same technique
we use to compute $\beta.$ The first time step to integrate (\ref{eq:190}) is
carried out with explicit Euler method, and from there on with the two-step,
second-order accurate, Leap-Frog method. We evaluate the integral in
(\ref{eq:mass_omega}) with the midpoint rule.

\section{Tests, Runs and Numerical Results}
\label{sec:numerical-results}

The numerical calculations we carry out in this work pursue two main objectives.
The first objective is to make plausible that the initial-boundary value
problem for (\ref{eq:hyp}),(\ref{eq:ell}) is well-posed. If we were simulating an
IBVP that is not well-posed, the expectation would be that almost any consistent
numerical simulation of the problem would fail to pass convergence tests,
numerical stability tests, or both. We show below that both kind of numerical
tests are passed satisfactorily by our numerical approximation. The second
objective is to study the behavior of the mass in these initial-boundary value
problems. In particular, we will show that for fixed initial data, the larger
the domain used in our calculation is, the longer and better the mass approaches
a constant value. 

We use in our runs two kinds of initial data which are smooth and strongly
decaying outside a small region (Gaussian functions). The first is
\begin{equation}\label{typical_idata}\begin{split}
\bar v_0(\rho,z)&= \exp\Bigl(\frac{(\rho-1/2)^2 + (z-L/2)^2}{0.1^2}\Bigr),\\
\bar v_{0t}(\rho, z)&= 0.
\end{split}
\end{equation}
which decays very fast as $(\rho,z)$ get away from $(1/2,L/2)$ and so
approximate very well a compact support data on the domains we use. The second
is the same kind of function but for $\bar v_{0t}$ instead of $\bar v_0.$ namely
\begin{equation}\label{typical_idata_2}\begin{split}
\bar v_0(\rho,z)&= 0, \\
\bar v_{0t}(\rho, z)&= 50 \exp\Bigl(\frac{(\rho-1/2)^2 + (z-L/2)^2}{0.1^2}\Bigr).
\end{split}
\end{equation}
Linearity of the problem tells us that the runs with $\bar v_0\neq 0,$ or $\bar
v_{0t}\neq 0,$ can be performed separately. A solution with general initial data
is the superposition of two solutions, one with each kind of data.

\begin{figure}[th]
\begin{center}
\includegraphics[width=8cm]{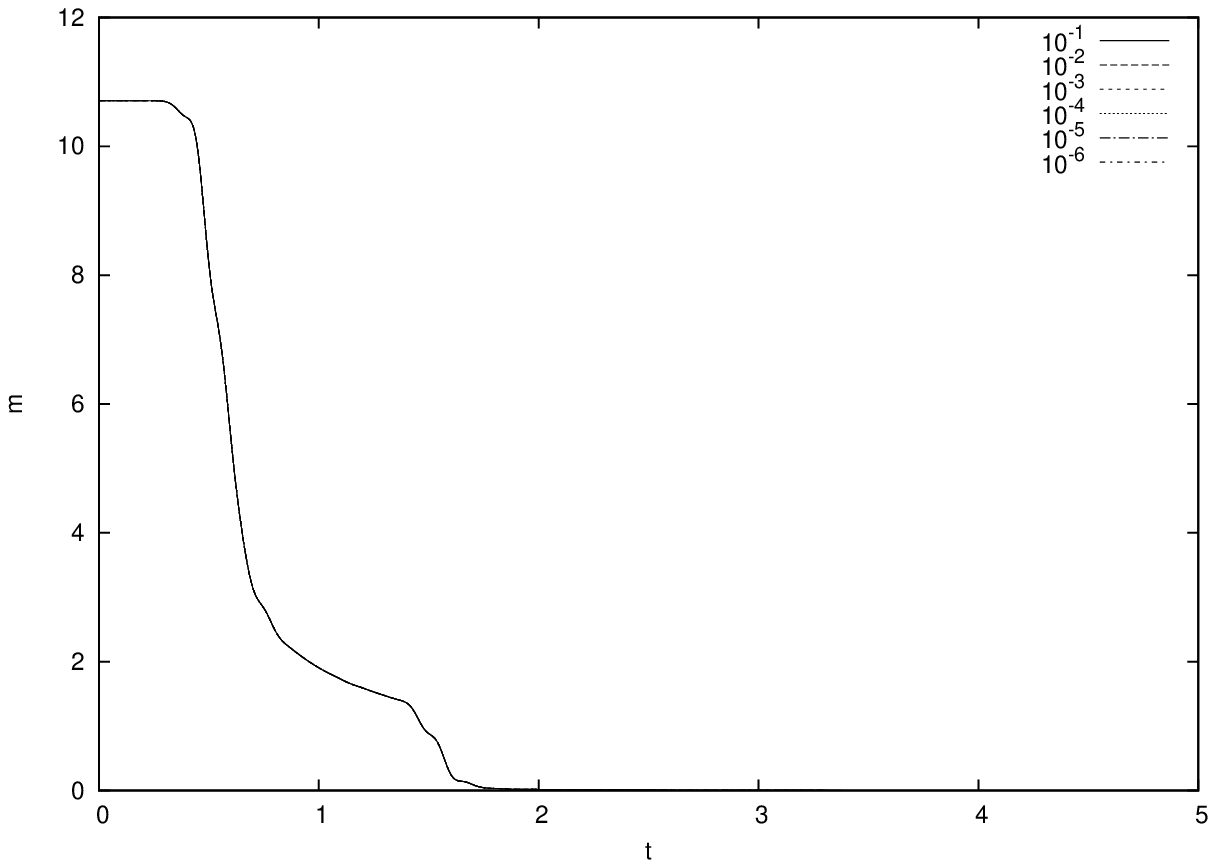}\\
\includegraphics[width=8cm]{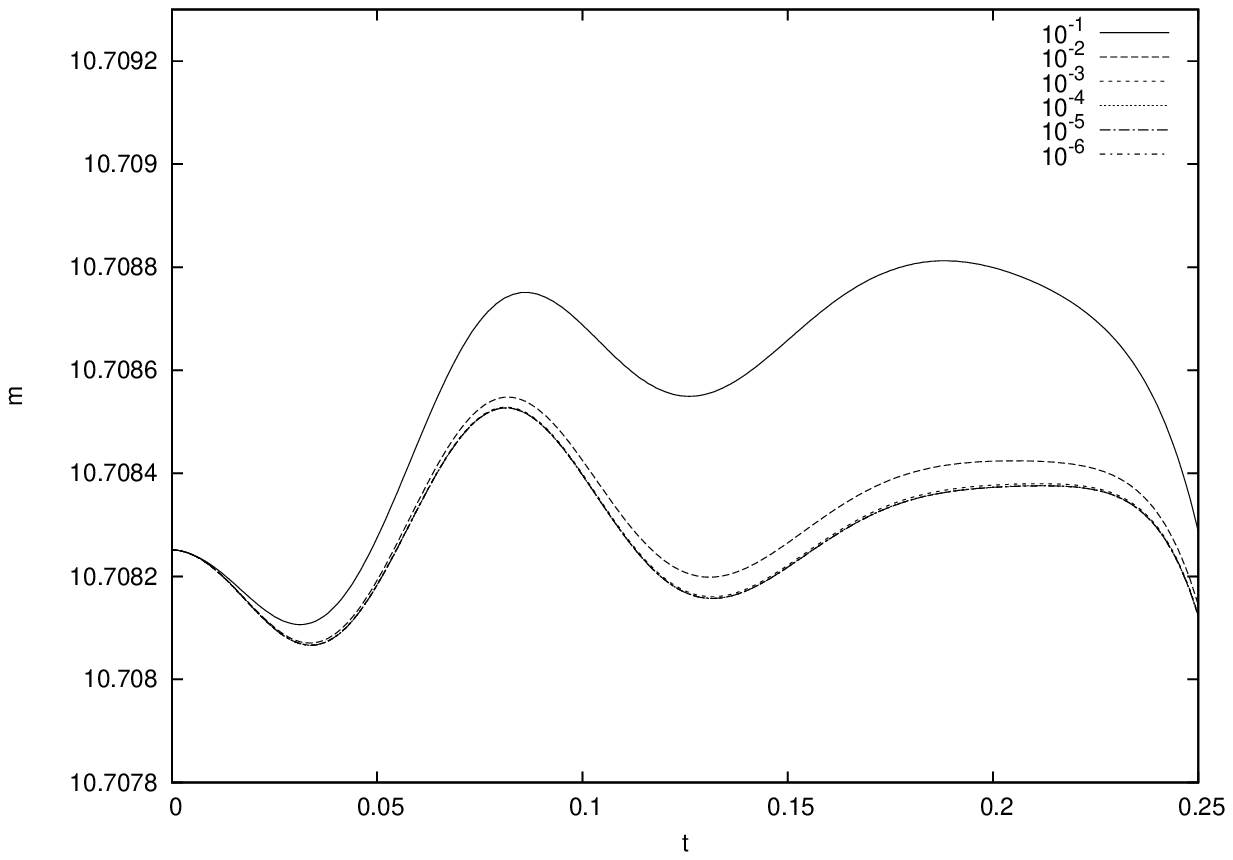}
\end{center}
\caption{Evolution of the mass for six different values of the tolerance
$\varepsilon.$ The upper plot shows the six runs in full mass scale. In this
scale the six curves look superimposed. The lower plot shows a detail of the
initial ``flat'' region in amplified scale.}\label{fig:epsilon_full}
\end{figure}

\paragraph{Elliptic Solver Tolerance.} We need to determine the value of
$\varepsilon$ to use in our runs.  To this end we perform runs for six different
values of $\varepsilon$ with all other parameters fixed to typical values in our
runs. In these tests runs we use initial data given by (\ref{typical_idata}) and
Sommerfeld boundary conditions (\ref{bc:axis}),(\ref{bc:som}). We then analyze
the different values of the mass obtained for the six solutions.  By comparing
the variations of $m_\Omega(t)$ with respect to the initial value of $m_\Omega,$ we see from
our runs show that the evolution is not very sensitive to the tolerance
$\varepsilon.$ Fig. \ref{fig:epsilon_full} shows that the plot for the different computed
masses superimpose when plotted in the full mass scale. The detail in the figure
shows convergence of $m_\Omega$ as $\varepsilon\to 0.$

In table \ref{table:epsilon} we show the maximum absolute difference between the
computed masses with respect to the most accurate one (corresponding to
$\varepsilon=10^{-6}$) and the time of occurrence.
\begin{table}[th]
\begin{tabular}{|c|c|c|c|}
\hline
$\varepsilon$ &  $\Delta m_\Omega$ &
$t_{max}$ & $\delta m_\Omega/m_{\Omega 0}$ \\
\hline
$10^{-1}$ & $1.27\times 10^{-3}$ & 0.390 & $1.19\times 10^{-4}$ \\
$10^{-2}$ & $1.81\times 10^{-4}$ & 0.406 & $1.69\times 10^{-5}$ \\
$10^{-3}$ & $1.45\times 10^{-5}$ & 0.481 & $1.36\times 10^{-6}$ \\
$10^{-4}$ & $1.08\times 10^{-6}$ & 0.503 & $1.01\times 10^{-7}$ \\
$10^{-5}$ & $8.19\times 10^{-8}$ & 0.509 & $7.65\times 10^{-9}$ \\
\hline
\end{tabular}
\caption{Values of $\Delta m_\Omega = \max_t|m_{\Omega\varepsilon}(t) -
m_{\Omega 10^{-6}}(t)|$ and time of occurrence for different values of the
tolerance $\varepsilon.$ Initial mass is $m_{\Omega
0}=10.7083.$}\label{table:epsilon}
\end{table}

Based on this test we choose to use $\varepsilon=10^{-3}$ in our further runs,
which gives more that necessary accuracy for our discussion (around $10^{-6}$
relative error.)

\paragraph{Convergence Tests.} To study convergence of the numerical solution we
perform two series of runs in a unitary square domain with the initial data
(\ref{typical_idata}). In the first series we use homogeneous Dirichlet boundary
conditions and in the second Sommerfeld boundary conditions.

Each series consists of four runs. In the successive runs we use $h=1/N$ with
$N=50; 100; 200; 400.$ In all runs $\delta t=h/10.$ Thus, in the second, third
and fourth runs both $h$ and $\delta t$ are divided by 2 with respect to the
previous run. Let us call $V^{(h)}(t)$ the solution computed using mesh-size
$h.$

\begin{figure}[ht]\includegraphics[width=8cm]{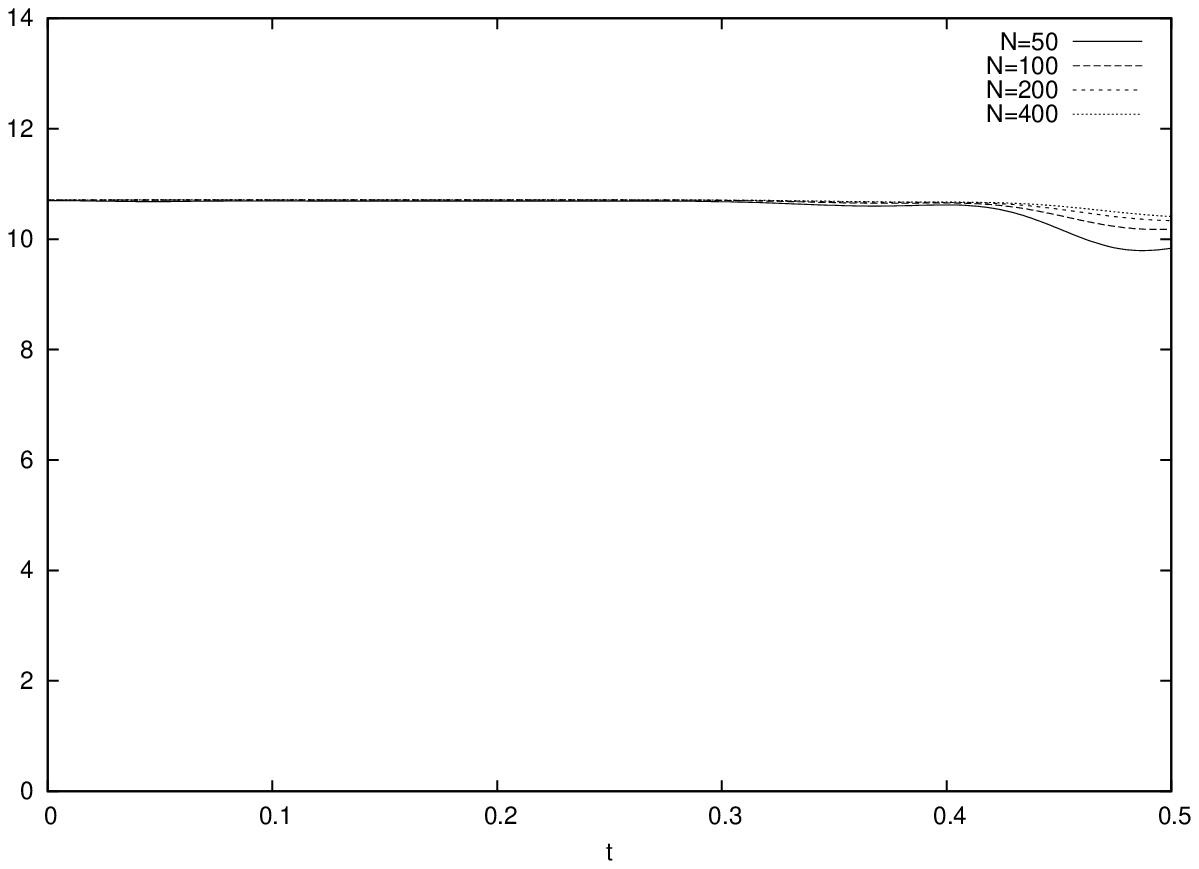}
\includegraphics[width=8cm]{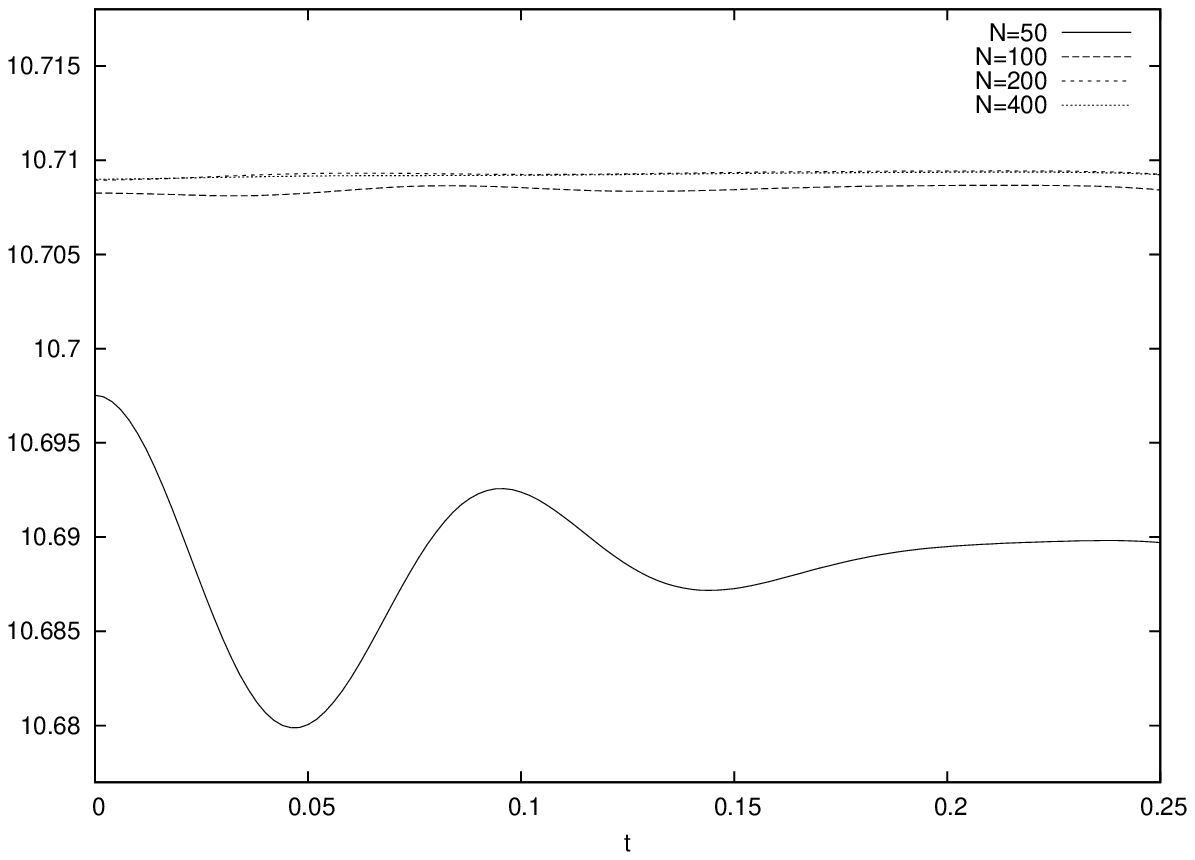}
\caption{Mass as function of time for evolution with homogeneous Dirichlet
boundary conditions. In the upper plot, in full mass scale, the
four curves look almost superimposed. In the lower plot a detail in expanded
mass scale shows that the curves converge to a limit curve when $h$ and
$\delta t$ diminishes}.\label{fig:dir_conv}
\end{figure}

\begin{figure}[ht]\includegraphics[width=8cm]{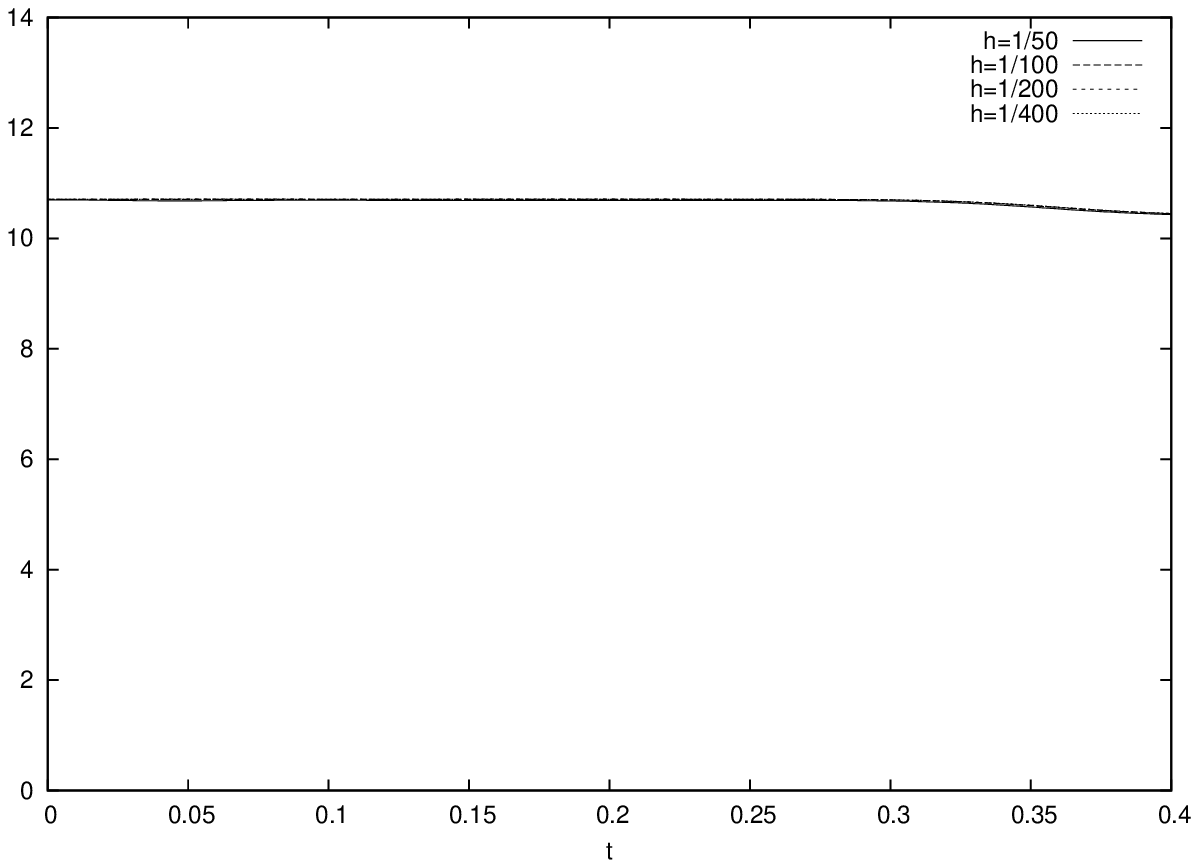}
\includegraphics[width=8cm]{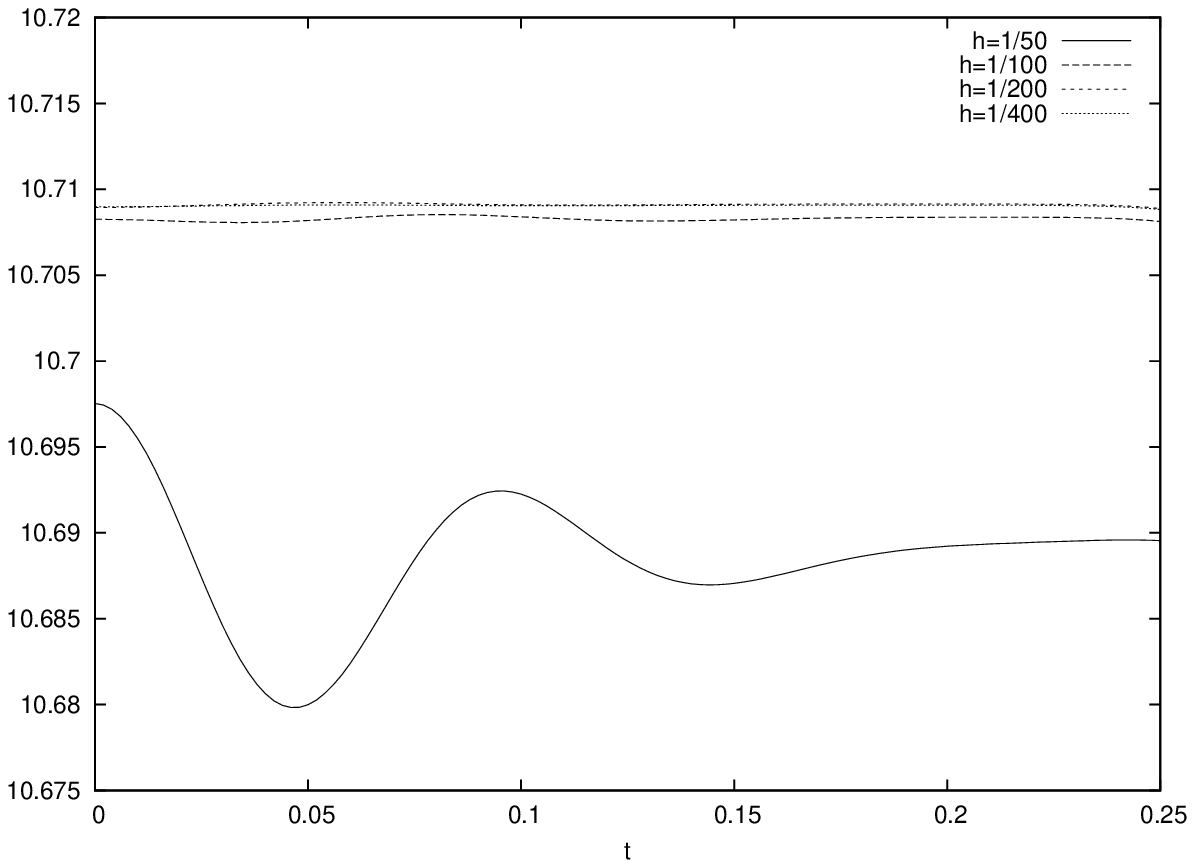}
\caption{Mass as function of time for evolution with Sommerfeld
boundary conditions. In the upper plot, in full mass scale, the
four curves look almost superimposed. In the lower plot a detail in expanded
mass scale shows that the curves converge to a limit curve when $h$ and
$\delta t$ diminishes}\label{fig:som_conv}
\end{figure}

The first, simplest and indirect, convergence test is to plot the
masses for each run as a function of time and check, graphically, whether they
converge as the value of $h$ diminishes. Figures \ref{fig:dir_conv} and
\ref{fig:som_conv} show that this is in fact the case.

A second more strict convergence and accuracy test is as follows.  We compute
the $L_2$ norm of the difference between two successive runs. A simple analysis
shows that, when the method is convergent and the mesh and time-step sizes are
small enough, the quotient
\begin{equation}\label{eq:Q}
Q_h(t_n)= \frac{\|V^{(h)}(t_n) - V^{(h/2)}(t_n)\|_{L_2}}{\|V^{(h/2)}(t_n) -
V^{(h/4)}(t_n)\|_{L_2}}
\end{equation}
approaches the value $2^p$ where $p$ is the accuracy order of the method. Our
method is fourth-order accurate in space and second-order in time. Therefore the
expectation is that we obtain values of $Q_h$ that are close to 4 at most times.

To compute the $L_2$-norms we use the midpoint rule to approximate the
integration on the coarsest grid of the two solutions being subtracted. Notice
that the coarse grid is not sub-grid of a the fine one, as they are displaced
from the domain boundaries by different amounts. Then, to evaluate the finest
solution on the coarse grid we need to interpolate this solution. To do this we
use bilinear interpolation.
 
\begin{table}[bth]
\begin{tabular}{|c|c|c|c|}
\hline
$t$ & $~\|V^{(2)} - V^{(3)}\|_{L_2}~$ & $\|V^{(3)} - V^{(4)}\|_{L_2}$ & $~Q_h(t)~$\\
\hline
0.05 &	5.2744$\times 10^{-5}$ &	1.3214$\times 10^{-5}$ &	3.9913\\
0.10 &	8.2053$\times 10^{-5}$ &	2.0665$\times 10^{-5}$ &	3.9706\\
0.15 &	9.2690$\times 10^{-5}$ &	2.3339$\times 10^{-5}$ &	3.9715\\
0.20 &	9.8158$\times 10^{-5}$ &	2.4938$\times 10^{-5}$ &	3.9360\\
0.25 &	1.1260$\times 10^{-4}$ &	2.8857$\times 10^{-5}$ &	3.9020\\
0.30 &	1.3325$\times 10^{-4}$ &	3.4668$\times 10^{-5}$ &	3.8436\\
0.35 &	1.6185$\times 10^{-4}$ &	4.4464$\times 10^{-5}$ &	3.6401\\
0.40 &	1.8421$\times 10^{-4}$ &	4.9874$\times 10^{-5}$ &	3.6934\\
0.45 &	2.2492$\times 10^{-4}$ &	6.9417$\times 10^{-5}$ &	3.2402\\
0.50 &	2.9719$\times 10^{-4}$ &	6.9240$\times 10^{-5}$ &	4.2923\\
\hline
\end{tabular}
\caption{Convergence and accuracy quotient for solutions with homogeneous
Dirichlet boundary condition. On the coarsest grid $h=10^{-2}.$}\label{table:dir_conv}
\end{table}
The results of this analysis are shown in the table \ref{table:dir_conv} and
\ref{table:som_conv}. The test is passed satisfactorily.

\begin{table}[bth]
\begin{tabular}{|c|c|c|c|}
\hline
$t$ & $~\|V^{(2)} - V^{(3)}\|_{L_2}~$ & $\|V^{(3)} - V^{(4)}\|_{L_2}$ & $~Q_h(t)~$\\
\hline
0.04 & 6.6922$\times 10^{-5}$ &	1.6703$\times 10^{-5}$ & 4.0067\\
0.08 & 6.2007$\times 10^{-5}$ &	1.5756$\times 10^{-5}$ & 3.9355\\
0.12 & 9.1495$\times 10^{-5}$ &	2.2958$\times 10^{-5}$ & 3.9854\\
0.16 & 9.2694$\times 10^{-5}$ &	2.3357$\times 10^{-5}$ & 3.9686\\
0.20 & 9.8226$\times 10^{-5}$ &	2.4911$\times 10^{-5}$ & 3.9431\\
0.24 & 1.0935$\times 10^{-4}$ &	2.7940$\times 10^{-5}$ & 3.9138\\
0.28 & 1.2407$\times 10^{-4}$ &	3.2083$\times 10^{-5}$ & 3.8671\\
0.32 & 1.4427$\times 10^{-4}$ &	3.8153$\times 10^{-5}$ & 3.7813\\
0.36 & 1.6740$\times 10^{-4}$ &	4.6440$\times 10^{-5}$ & 3.6047\\
0.40 & 1.8389$\times 10^{-4}$ &	5.0471$\times 10^{-5}$ & 3.6434\\
\hline
\end{tabular}
\caption{Convergence and accuracy quotient for solutions with Sommerfeld
boundary condition. On the coarsest grid $h=10^{-2}.$}\label{table:som_conv}
\end{table}

\paragraph{Stability Tests.} Numerical stability means that the solution to the
IBVP stays bounded during time evolution. Typical signs of instability are the
appearance of artifacts in the plot of the solution as a consequence of
evolution and in most cases, after a while, the complete break-down of the
solution. If an instability has its root on the ill-posedness of the analytic
problem underneath, the expectation is that some high frequency modes of the
solution explode exponentially fast and are detected at very short times of the
numerical evolution. For some more benign ill-posed problems (like
weakly-hyperbolic problems) the growing of instabilities is only polynomial and
it may take longer to detect them.

\begin{figure}
\begin{center}
\includegraphics[width=8cm]{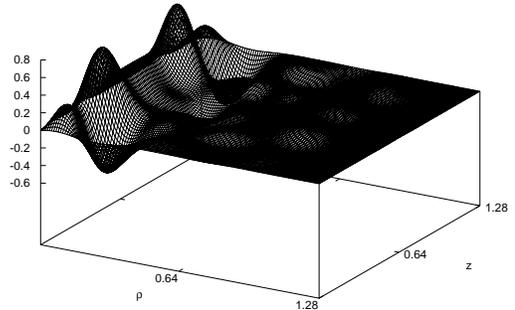}\\
\includegraphics[width=8cm]{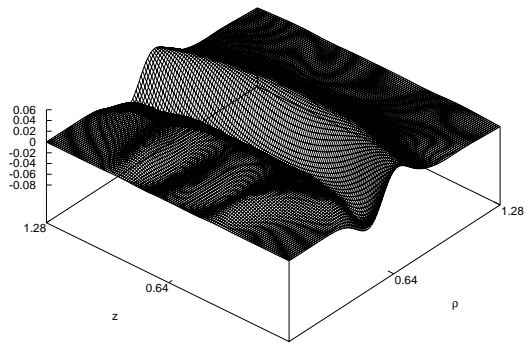}
\end{center}
\caption{Plots of the solution $\bar v(t).$ Both plots of solutions computed on
  a grid with $128\times 128$ gridpoints and initial data given by
  (\ref{typical_idata}). Upper plot is the solution with homogeneous Dirichlet
  boundary conditions at time $t=3.0,$ while lower plot is the solution with
  Sommerfeld boundary conditions at time $t=1.25.$}\label{fig:sol}
\end{figure}

We performed several series of runs using both kinds of boundary conditions
(\ref{bc:axis}),(\ref{bc:dir}) or (\ref{bc:axis}),(\ref{bc:som}) and both kinds
of initial data (\ref{typical_idata}) or (\ref{typical_idata_2}) on different
domains and during several time intervals. We studied
the plots of the solutions in all cases and they always look smooth, agreement
with the boundary conditions imposed and never showed any sort of strange
artifact. Typical plots for $\bar v(\rho,z,t)$ are shown in Fig. \ref{fig:sol}.
We have also studied the plots of $\beta(\rho,z,t)$ in these runs and no sign
of instability showed.

A second, physically meaningful, test for stability is provided by the study of
the mass $m_\Omega$ which in this problem is a sort of incomplete $H^2$ Sobolev
norm of the solution. As explained in Sec. \ref{sec:prop-line-equat} the
mass is conserved for the Cauchy problem in the whole space. On bounded domains
this is no longer true, but we expect that it stays bounded when using
homogeneous Dirichlet boundary conditions, and that it goes to zero when using
Sommerfeld boundary conditions.  We analyze the behavior of the mass below.

\paragraph{Behavior of the Mass.} As explained before the mass, defined by (\ref{eq:100}) and
(\ref{eq:mass_omega}), is a conserved quantity when the Cauchy problem is
considered in the whole space (i.e., $\Omega$ is ${\mathbb R}^2_+$). In our
numerical tests we solve the initial boundary value problem on compact domains
where no known boundary conditions imply mass conservation. However, an
interesting study for the mass evolution can be done as follows. We solve the
IBVP on domains of different size but use, in all runs, the same initial data,
at the same distance from the symmetry axis. The initial data are chosen to
decay exponentially fast outside a region which is small compared to the
smallest of the domains we use.  Clearly, the expectation is that the larger the
domain is the closest to constant the mass stays during evolution.

\begin{figure}
\begin{center}
\includegraphics[width=8cm]{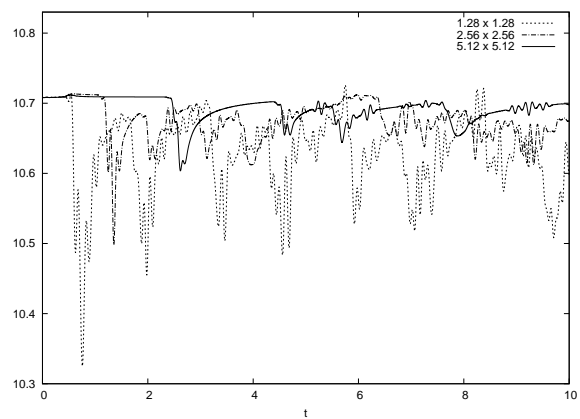}
\end{center}
\caption{Evolution of the mass $m_\Omega$ as a function of time for three
  solutions with homogeneous Dirichlet boundary conditions and the same initial
  data but on domains of different size. In the upper right corner the each
  curve is associated to the corresponding domain.}\label{fig:dir_domains}
\end{figure}

\begin{figure}
\begin{center}
\includegraphics[width=8cm]{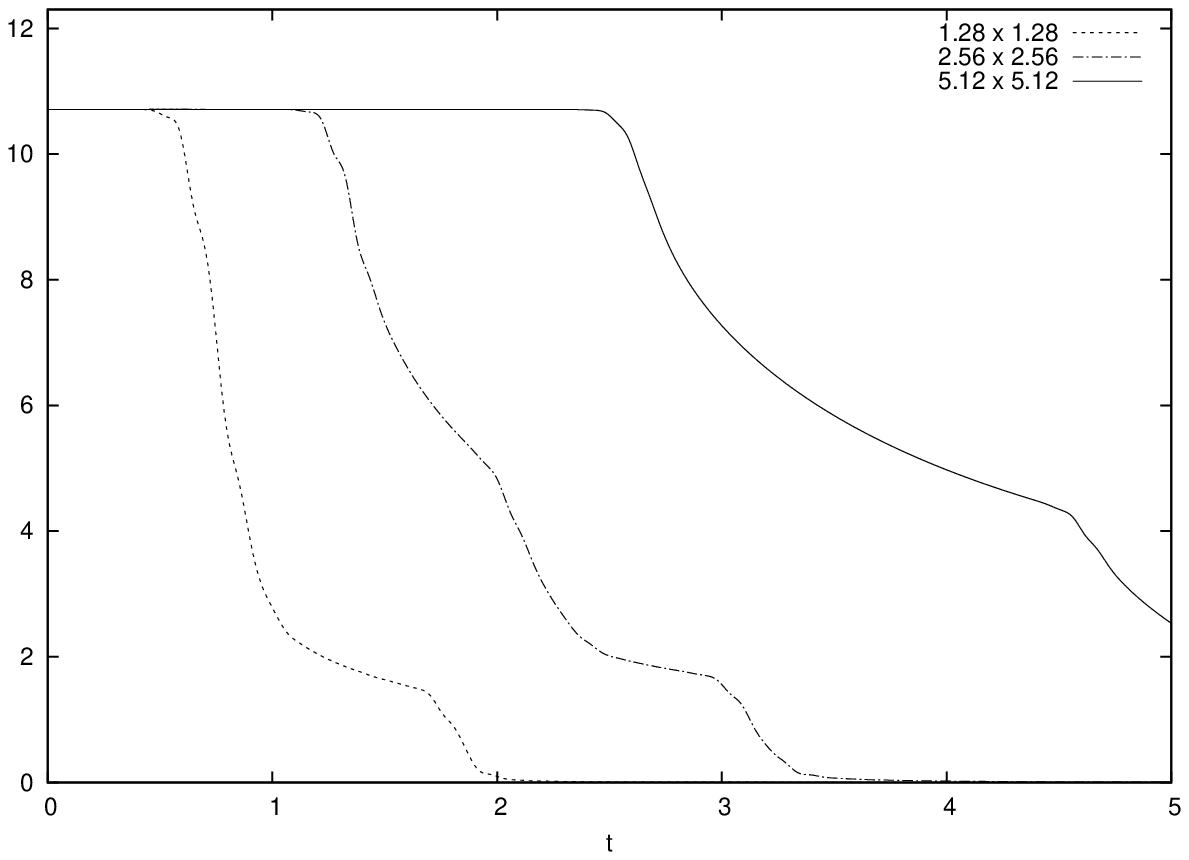}
\includegraphics[width=8cm]{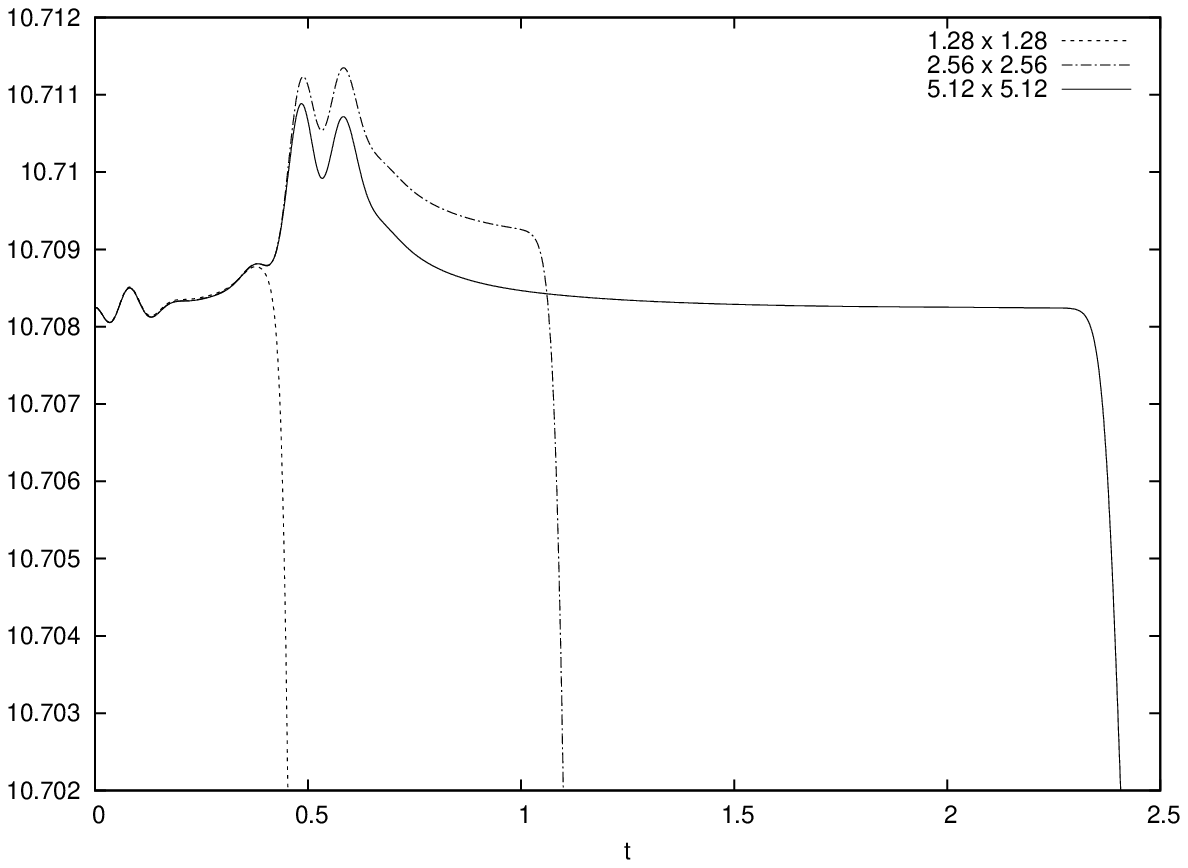}
\end{center}
\caption{Evolution of the mass $m_\Omega$ as a function of time for three
  solutions with Sommerfeld boundary condition, the same initial data but on
  domains of different size. In the upper right corner the each curve is
  associated to the corresponding domain. The lower plot shows in amplified
  scale that the ``flat'' region presents very small variations of around
  $0.03\%.$}\label{fig:som_domains}
\end{figure}

We do series of runs for homogeneous Dirichlet boundary conditions and for
Sommerfeld (outgoing waves) boundary conditions. The plots for the Dirichlet
case are shown in Fig. \ref{fig:dir_domains}. Observe that the plot is not on
full mass scale. The three curves show an almost constant initial region and
then variations of small relative amplitude. After an initial peak immediately
after the constant region the amplitude of the variations is, roughly speaking,
$2\%$ for the $1.28\times1.28$ domain, $1\%$ for the $2.56\times2.56$ domain
and $0.6\%$ for the $5.12\times5.12$ domain. The amplitude clearly diminishes
when the domain increases size.
For the case of Sommerfeld boundary conditions, the plots of the mass evolution
can be seen in Fig. \ref{fig:som_domains}. This series of three runs is
totally analogous to the previous case. The only change is the boundary
condition used.
As can be inferred from the plot in full mass scale, the energy leaks though the
boundary as expected.


\section{Final comments}
\label{sec:final-comments}

In this article we have deduced the linear system
\eqref{eq:119b} and \eqref{eq:120b} and we have analyzed some of its
properties. Among them, the most relevant are the mass conservation and the
numerical stability. The main open problem is to prove that this system is
well-posed. Remarkable enought, it seems to be not much literature on this class of
linear systems which are singular at the axis.

Once the well-posedness problem is solved, we believe that the future research on
the subject can be divided in two paralel but complementary roads.  The first
one is to extend the well-posedness from the linear system to the full Einstein
equations in the maximal-isothermal gauge. The non-linear lower order terms
introduce extra difficulties (see \cite{Rinne:2008tk}). There are many possible
evolutions schemes (see the discussion in \cite{Rinne:thesis}).  It is very
likely that few of them (or may be only one) are well-posed.  If this is the
case, the resolution of the well-posedness question will lead us to select (or even
discover) the correct evolution scheme.  After the local problem is solved,
 the next step is to use the global conservation of the mass to control the
full non-linear evolution in this gauge. A natural first example would be to
recover the non-linear stability of Minkowski \cite{Christodoulou93} in this
gauge. The expectation is that the mass formula will provide a simpler (and
different) kind of approach to this problem; although, of course, always
resticted to axial symmetry. The ultimate and difficult goal is to say
something, in this gauge, about the non-linear stability of a black hole in
axial symmetry.

The second road is the study axially symmetric perturbation but with a black
hole as background solution.  Linear stability of the Kerr black hole is a
relevant open problem which is currently intensively studied (see the review
articles \cite{Finster:2008bg}, \cite{Dafermos:2008en} and references therein).
The expectation is that the mass formula can help to prove linear
stability under axially symmetric perturbation of the Kerr black hole.

\begin{acknowledgments} 
   S. D. thanks Piotr Chru\'sciel and Helmut Friedrich for useful
   discussions. These discussions took place at the Mathematisches
   Forschungsinstitut Oberwolfach during the workshop ``Mathematical Aspects of
   General Relativity'', October 11th -- October 17th, 2009.  S. D. thanks
   Andr\'es Ace\~na for useful discussions that took place at the
   Max-Planck-Institut f\"ur Gravitationsphysik (Albert-Einstein-Institut)
   during the conference ``Space, Time and Beyond'', October 19th -- October
   21th, 2009.  S. D. thanks the organizers of these events for the invitation
   and the hospitality and support of the Mathematisches Forschungsinstitut
   Oberwolfach and the Max-Planck-Institut f\"ur Gravitationsphysik
   (Albert-Einstein-Institut).

   S. D. is supported by CONICET (Argentina).  This work was supported in part
   by grant PIP 6354/05 of CONICET (Argentina), grant 05/B415 Secyt-UNC
   (Argentina) and the Partner Group grant of the Max Planck Institute for
   Gravitational Physics, Albert-Einstein-Institute (Germany).
 \end{acknowledgments} 

\appendix

\section{Useful formulas}
We collect in this appendix some useful formula that are used in the main part
of this article. The conformal Killing operator in 2-dimensions with respect to
the  metric $q_{AB}$ is defined by 
\begin{equation}
  \label{eq:43}
 (\ckq \beta )_{AB} =D_A\beta_B + D_B\beta_A -q_{AB}  D_C\beta^C. 
\end{equation}
For the particular case of a flat metric $\delta_{AB}$ this definition reduce
to 
\begin{equation}
  \label{eq:67}
  (\ck \beta )_{AB} =\partial_A\beta_B + \partial_B\beta_A -\delta_{AB}  \partial_C\beta^C. 
\end{equation}
For this operator we have the following identity often used in the article
\begin{equation}
  \label{eq:28}
  \partial^B (\ck \beta )_{AB}=\Ld \beta_A.
\end{equation}

The Christofell symbols of the metric $q_{AB}$ defined by \eqref{eq:10} are given by
\begin{equation}
  \label{eq:87}
  \Gamma^C_{AB}= \delta^C_B \partial_A u +\delta^C_A \partial_B u-\partial^Cu
  \delta_{AB},  
\end{equation}
and the  Ricci tensor is given by
\begin{equation}
  \label{eq:11}
 \rd_{AB}=-\Ld u\, \delta_{AB}, \quad \rd=-2e^{-2u}\Ld u.
\end{equation}

Under the conformal rescaling \eqref{eq:10} the diferential operators relevant
in this article transform as follows 
\begin{align}
  \label{eq:64}
 \Lq f &=e^{-2u} \Ld f,\\
\label{eq:151}
\ckq(\beta)_{AB}& =e^{2u}\ck(\hat\beta)_{AB},\\
\label{eq:152}
D_B\chi^{AB}& = e^{-4u} \partial_B \hat \chi^{AB}, 
\end{align}
where we have defined 
\begin{equation}
  \label{eq:141}
  \beta_A=e^{2u}\hat \beta_A \quad \chi^{AB}=e^{-4u}\hat \chi^{AB}.
\end{equation}
We follow the convention that the indices for hat quantities are moved with the
flat metric $\delta_{AB}$ and inidices of non-hat quantities with the metric
$q_{AB}$. Then,  we have 
\begin{equation}
  \label{eq:45b}
  \chi_{AB}=\hat\chi_{AB}, \quad \beta^A=\hat \beta^A. 
\end{equation}
That is why we suppress the hat notation for the tensors $\hat\chi_{AB}$ and
$\hat \beta^A$ in the main part of this article.

Take an arbitray spacelike foliation on $\N ,h_{ab}$. The $2+1$ decomposition
of the wave operator is given by
\begin{equation}
  \label{eq:68}
  \Box f = -f'' + \Lq f + D_A f \frac{D^A \alpha}{\alpha} + f' \chi, 
\end{equation}
where have made use of the following useful formulas
\begin{equation}
  \label{eq:52}
  n^a\nabla_a n_A = \frac{\partial_A \alpha}{\alpha}, \quad  n^a\nabla_a n_t =
  \frac{\beta^A\partial_A \alpha}{\alpha}.
\end{equation}


\end{document}